\documentclass[12pt]{article}
\usepackage{amssymb,amsmath}
\usepackage{amsfonts}
\usepackage{eufrak}
\usepackage{graphicx}
\usepackage{amsbsy}
\usepackage{mathrsfs}

\DeclareMathAlphabet {\mathitbf}{OML}{cmm}{b}{it}

\begin{document}

\title{Dispersion Relations for Cold Plasmas around \\
Reissner-Nordstr\"{o}m Black Holes}

\author{M. Hossain Ali \thanks{E-mail: m$\_$hossain$\_$ ali$\_$bd@yahoo.com (corresponding author)}\;
and
M. Khayrul Hasan \thanks
{Department of Mathematics, Shahjalal University of Science and Technology, Sylhet-3100, Bangladesh. E-mail: khayrulmat@gmail.com}\\
{\it Department of Applied Mathematics,}\\
{\it University of Rajshahi, Rajshahi-6205, Bangladesh}}

\date{}
\maketitle

\begin{abstract}
We investigate the general relativistic magnetohydrodynamic  (GRMHD) equations for cold plasma around the Reissner-Nordstr\"{o}m black hole. Applying 3+1 spacetime split we linearize the perturbed equations for non-magnetized/magnetized plasma in both rotating and non-rotating background. By Fourier analyze we then derive dispersion relations and investigate the existence of waves with positive angular frequency in the vicinity of the black hole horizon. The analysis finds propagation of negative phase and group velocities for rotating magnetized surroundings.
\end{abstract}

\noindent
\textbf {PACS number(s)}: 95.30.Sf, 95.30.Qd, 04.30.Nk

\newpage

\section{Introduction}\label{Intro}

Black holes, the most simple predictions of the general relativity, are one of the most enigmatic constructs in the present day physics. They draw physicists' attention as the paradigmatic objects to test possible quantum theories of gravity. Although still there is no convincing observational data in favor of conclusively proving the existence of black holes in the universe, there exist certainly sufficient evidences that make the study of such objects and the effects on their environment a matter of great importance to astrophysics. However, black holes are not objects of direct observing, so we must have to observe them indirectly through the effects they exert on their environment. They will greatly affect the surrounding plasma medium (which is highly magnetized) with their enormous gravitational fields, and hence, plasma physics in the vicinity of a black hole has become a subject of obvious interest in astrophysics. In the immediate vicinity of a black hole general relativity applies. It is therefore of interest to formulate plasma physics problems in the context of general relativity.

The energy flux carried out by black holes (or compact stars) produces a relatively large magnetic field. The study of stationary configurations and dynamic evolution of conducting fluid in a magnetosphere of massive black holes demands the theory of general relativistic magnetohydrodynamics (GRMHD). The equations of GRMHD theory describe the aspects of interaction of relativistic gravity with plasma's magnetic field. The study of plasmas in the black hole environment is important because a successful study of waves will be of great value in aiding the observational identification of black hole candidates.

An isolated black hole can have an electromagnetic field, if it is endowed with a net electric charge \cite{one,two,three,four}. Since a collapsed object can have a very strong effect on an electromagnetic field, it is of interest to determine this effect using GRMHD equations when a black hole is placed in an external electromagnetic field.

A covariant formulation of the theory based on the fluid equations in curved spacetime has so far proved unproductive because of the curvature of four-dimensional spacetime in the region surrounding a black hole. The 3+1 formulation of general relativity, developed by Thorne, et al \cite{five,six,seven}, provides a method in which the electromagnetic equations and the plasma physics at least look somewhat similar to the usual formulations in flat spacetime while taking accurate account of general relativistic effects such as curvature.

In 3+1 formalism of Thorne, Price, and Macdonald (TPM) \cite{eight}, work connected with black holes has been facilitated by the replacement of the hole's event horizon with a membrane endowed with electric charge, electrical conductivity, and finite temperature and entropy. Mathematically the membrane paradigm is analogous to the standard, full general relativistic theory of black holes so far as physics outside the event horizon is concerned, and moreover, the formulation of all physics in this region turns out to be very much simpler than it would be using the standard covariant approach of general relativity.

The pioneer work of 3+1 spacetime split, called ADM (Arnowitt, Deser, and Misner) formalism, was done in 1962 \cite{nine} to study the quantization of the gravitational field. Since then, their formulation has most been applied in studying numerical relativity \cite{ten}. TPM extended the ADM formalism to include electromagnetism and applied it to study electromagnetic effects near the Kerr black hole. As a result, their work has opened up many possibilities for studying electromagnetic effects on plasmas in the black hole environment.

In recent years there have been attempts to exploit the 3+1 formalism. Zhang \cite{eleven,twelve} considered the case of perfect GRMHD waves in the vicinity of Kerr black hole and discussed the linearized waves for the cold (negligible particle pressure) plasma propagating in two-dimensions. Holcomb and Tajima \cite{thirteen}, Holcomb \cite{fourteen}, and Dettmann et al. \cite{fifteen} investigated some properties of wave propagation in the Friedmann universe. Khanna \cite{sixteen} derived the GRMHD equations for two-fluid plasma in Kerr black hole. Ant\'{o}n et al. \cite{seventeen} investigated various test simulations and discussed magneto-rotational instability of accretion disks. Anile \cite{eighteen} worked on relativistic shocks/simple waves in magneto-fluids in cold relativistic plasma. Komissarov \cite{nineteen} discussed the Blandford-Znajek monopole solution in black hole electrodynamics. Buzzi et al. \cite{twenty,twenty one} described a general relativistic version of two-fluid plasma physics in TPM formulation and developed a linearized treatment of plasma waves in analogy with the special relativistic formulation of Sakai and Kawata \cite{twenty two}. They also investigated the one dimensional radial propagation of transverse and longitudinal waves near the Schwarzschild black hole.

Recently, Sharif and Sheikh \cite{twenty three} investigated cold plasma in the vicinity of the Schwarzschild black hole horizon by using $3+1$ formalism of the GRMHD equations and described the dispersion relation.

In this paper we apply TPM formalism of the GRMHD equations to study the dynamical magnetosphere of the Reissner-Nordstr\"{o}m (RN) spacetime, which is the solution of the coupled Einstein-Maxwell equations and describes a spherically symmetric black hole endowed with electric or magnetic charge. The RN solution is important due to its intrinsic interest \cite{twenty four,twenty five} specifically in the context of its recent applications in Hawking radiation \cite{twenty six,twenty seven,twenty eight,twenty nine}. If the horizon is sufficiently small, the magnetically charged RN solution develops a classical instability in the context of spontaneously broken gauge theories, which has significant implications for the evolution of a magnetically charged black hole \cite{thirty}. It leads, in particular, to the possibility of evaporating a black hole completely, leaving in its place a nonsingular magnetic monopole. The magnetic monopole hypothesis was propounded by Dirac \cite{thirty one,thirty two} relatively long ago. The ingenious suggestion by Dirac that magnetic monopole does exist in nature was neglected due to the failure to detect such a particle. However, in recent years the development of gauge theories has shed new light on it. Moreover, in asymptotically flat space, the extremal RN black holes hold an important and controversial status in black hole physics. Usually, these black holes were supposed to be a limiting case of non-extremal black holes \cite{thirty three}. This conventional view was challenged in \cite{thirty four} on the basis of the fact that the topology of the extremal and non-extremal black holes have qualitative differences. On the basis of these differences, it was argued that extremal RN black holes have zero entropy with no definite temperature in spite of having a nonzero horizon area \cite{thirty four,thirty five}. The extremal RN black holes are also important in the context of supergravity theories \cite{thirty six,thirty seven,thirty eight}. As shown in \cite{thirty nine}, an exact solution of these black holes exists in (super)string theory. Thus aspects of the RN solution must be of interest in a broader contest. In view of this reason, our study of cold plasma in the environment close to the event horizon of the RN black hole is interesting. The result we have obtained reduces to that of the Schwarzschild black hole \cite{twenty three} when the charge term vanishes.

This paper is arranged as follows. In Section 2, we summarize the GRMHD equations in the RN black hole magnetosphere in 3+1 formalism. We investigate the GRMHD equations for cold plasma in the case of rotating magnetized background in Section 3. In Section 4, we present our study of non-magnetized plasma in rotating background. We investigate the dispersion relations for the non-rotating background in Section 5. Finally, in Section 6 we present our remarks. We use natural units: $G=c=1$.

\section{3+1 Formulation of GRMHD Equations around RN Black Holes}\label{sec2}

The 3+1 formulation of general relativity is based on the concept of selecting a preferred set of spacelike hypersurfaces which form the level surfaces of a congruence of timelike curves. The choice of a particular set of these hypersurfaces constitutes a time slicing of spacetime. The hypersurfaces considered here are of constant universal time $t$.

In TPM notation, the metric of the Reissner-Nordstr\"{o}m (RN) black hole with magnetic charges \cite{fourty,fourty one} is given by
\begin{eqnarray}
&&ds^2=g_{\mu \nu }dx^\mu dx^\nu =-fdt^2+\frac{dr^2}{f}+r^2(d\theta ^2+{\rm sin}^2\theta d\varphi ^2),\nonumber\\
&&f(r)=1-\frac{2M}{r}+\frac{Q^2}{r^2},\qquad Q^2=Q_e^2+Q_m^2\label{eq1}.
\end{eqnarray}
The components $x^\mu $ denote spacetime coordinates and $\mu , \nu =0, 1, 2, 3$. The vector potential of the dually charged RN black hole has non-vanishing components: $A_t=Q_e/r$, $A_\varphi =-Q_m{\rm cos}\theta $, with corresponding field strength: $F_{01}=Q_e/r^2$, $F_{23}=Q_m/r^2$, describing electric charge $Q_e$ and magnetic charge $Q_m$.

The metric (\ref{eq1}) describes: (i) the Schwarzschild black hole for $Q=0$; (ii) the magnetically charged Reissner-Nordstr\"{o}m black hole \cite{thirty} for $Q_e=0$, $0<Q_m<M$; (iii) the generic Reissner-Nordstr\"{o}m black hole for $Q_m=0$, $0<Q_e<M$; and (iv) the extremal Reissner-Nordstr\"{o}m black hole for $Q=M$.

According to the null hypersurface equation, the metric function $f(r)$ can yield an event horizon at $r=r_+$ and an Cauchy (inner) horizon at $r=r_-$, where
\begin{equation}
r_\pm =M\pm \sqrt{M^2-Q^2}\label{eq2}.
\end{equation}
The curvature singularity at $r=0$ is hidden behind these horizons for $Q<M$, while for $Q>M$ the metric (\ref{eq1}) does not describe a black hole at all but rather a physically forbidden naked singularity, as no real value is found in (\ref{eq2}). The mass parameter $M$ of the RN solution is inseparably connected with the charge parameter $Q$ \cite{fourty two}. This means that $M=0$ only when $Q=0$.

When $Q=0$, $r_+=2M$ is the usual Schwarzschild horizon. In the extremal case of $Q=M$, the two values $r_\pm$ coincide and the horizon becomes degenerate. The extremal spacetime is just the flat space vacuum for $Q=0$. An $M>Q$ black hole will tend to Hawking radiate down to its extremal $M=Q$ state. The Hawking temperature \cite{fourty three}
\begin{equation}
T_H=\frac{\sqrt{M^2-Q^2}}{2\pi(M+\sqrt{M^2-Q^2})^2}\label{eq3}
\end{equation}
vanishes as the black hole attains its extremal limit. Thus the extremal black hole is typically quantum mechanically as well as classically a stable object. However, this is not true for extremal electrically charged black holes in our world \cite{fourty four}. The extremal black hole is quantum ground state of the charge $Q$ superselection sector of the Hilbert space. This view is reinforced by the $N=2$ supersymmetric version of the theory.

The hypersurfaces of constant universal time $t$ define an absolute three-dimensional space described by the metric $ds^2=g_{ij}dx^idx^j$, where the indices $i,j,$ refer to coordinates in absolute space and range over 1, 2, 3. The fiducial observers (FIDOs), i.e. the observers remaining at rest with respect to this absolute space, measure their proper time $\tau$ using clocks that they carry with them and make local measurements of physical quantities. Then all their measured quantities, such as velocities $\textbf{V}$ and fields $\textbf{B}$ and $\textbf{E}$ are defined as FIDO locally measured quantities and all rates as measured by the FIDOs are measured using FIDO proper time. In making these measurements the FIDOs use a local Cartesian coordinate system that has basis vectors of unit length tangent to the coordinate lines:
\begin{equation}
{\bf e}_{\hat r}=\sqrt{f}\frac{\partial }{\partial r},\hspace{1.2cm}{\bf e}_{\hat \theta }=\frac{1}{r}\frac{\partial }{\partial \theta },\hspace{1.2cm}{\bf e}_{\hat \varphi }=\frac{1}{r{\rm sin}\theta }\frac{\partial }{\partial \varphi }\label{eq4}.
\end{equation}

The ratio of the rate of FIDO proper time to that of universal time is defined in terms of redshift factor:
\begin{equation}
\alpha (r)\equiv \frac{d\tau }{dt}=\sqrt{1-\frac{2M}{r}+\frac{Q^2}{r^2}}\label{eq5},
\end{equation}
called the lapse function. It measures the amount of FIDO proper time elapsed during the passage of a unit amount of universal time.

If a spacetime viewpoint is considered rather than a 3+1 split of spacetime, the set of orthonormal vectors also includes the basis vector for the time coordinate given by
\begin{equation}
{\bf e}_{\hat 0}=\frac{d}{d\tau}=\frac{1}{\alpha }\frac{\partial }{\partial t}\label{eq6}.
\end{equation}
The FIDO proper time $\tau$ acts as a local laboratory time, where the FIDOs have the role of \lq \lq local laboratories\rq \rq . It does not provide a slicing of spacetime as it is not a global coordinate. The RN time coordinate $t$ is the logical choice to satisfy this role and in fact slices spacetime in the way that the FIDOs would do physically. All subsequent equations are therefore expressed in terms of the universal time coordinate $t$ rather than the FIDO proper time $\tau$.

The lapse function $\alpha$ acts as a gravitational potential and governs the ticking rates of clocks and redshifts as well. It also measures the gravitational acceleration felt by a FIDO:
\begin{equation}
\textbf{a}=\nabla\ln\alpha=\frac{1}{\alpha }\left(\frac{M}{r^2}-\frac{Q^2}{r^3}\right){\bf e}_{\hat r}\label{eq7}.
\end{equation}
The rate of change of any scalar physical quantity or any three-dimensional vector or tensor, as measured by a FIDO, is defined by the convective derivative
\begin{equation}
\frac{D}{D\tau }\equiv \left(\frac{1}{\alpha }\frac{\partial }{\partial t}+{\bf V}\cdot \nabla \right)\label{eq8},
\end{equation}
where $\bf V$ is the velocity of a fluid as measured locally by a FIDO.

The Rindler coordinate system, in which space is locally Cartesian, provides a good approximation to the RN metric near the event horizon. However, the essential features of the horizon and the 3+1 split are retained without the complication of explicitly curved spatial three-geometries. The RN metric is approximated in Rindler coordinates by
\begin{equation}
ds^2=-\alpha^2dt^2+dx^2+dy^2+dz^2,\label{eq9}
\end{equation}
where
\begin{equation}
x=r_+\left(\theta -\frac{\pi }{2}\right),\qquad y=r_+\varphi ,\qquad z=2r_+\alpha\label{eq10}.
\end{equation}
The standard lapse function is again denoted by $\alpha$ and simplifies in Rindler coordinates to $z/(2r_+)$ , where $r_+$ is the location of the event horizon of the RN black hole. This function vanishes at the horizon which we can place at $z=0$ and it increases monotonically as $z$ increases from $0$ to $\infty $.

Maxwell's equations in 3+1 formalism take the following form:
\begin{eqnarray}
\nabla \cdot {\bf B}&=&0,\label{eq11}\\
\nabla \cdot {\bf E}&=&4\pi \rho_e,\label{eq12}\\
\frac{\partial {\bf B}}{\partial t}&=&-\nabla \times (\alpha {\bf E}),\label{eq13}\\
\frac{\partial {\bf E}}{\partial t}&=&\nabla \times (\alpha {\bf B})-4\pi \alpha {\bf j},\label{eq14}
\end{eqnarray}
where $\rho_e$ and $\bf j$ are electric charge and current density, respectively. For the perfect MHD (i.e., MHD with perfectly conducting) assumption there exists no electric field in the fluid's rest frame, i.e., ${\bf E} + {\bf V}\times {\bf B} =0$. Under this condition the equation for the evolution of magnetic field (\ref{eq13}) becomes
\begin{eqnarray}
\frac{\partial{\bf B}}{\partial t}=\nabla \times(\alpha{\bf V\times{\bf B}})\label{eq15}
\end{eqnarray}
The conservation of mass and momentum equations are written, respectively, as follows \cite{eleven}:
\begin{eqnarray}
&&\frac{\partial(\rho_o\mu)}{\partial t}+\{(\alpha {\bf V})\cdot\nabla\}(\rho_o\mu)\nonumber\\
&&+\rho_o\mu\gamma^2{\bf V}\cdot\frac{\partial{\bf V}}{\partial t}+\rho_o\mu\gamma^2{\bf V}\cdot(\alpha{\bf V}\cdot\nabla){\bf V}+\rho_o\mu\{\nabla\cdot(\alpha{\bf V})\}=0\label{eq16},
\end{eqnarray}
\begin{eqnarray}
\left\{\left(\rho_o\mu\gamma^2 +\frac{{\bf B}^2}{4\pi}\right)\delta_{ij}+\rho_o\mu\gamma^4V_iV_j-\frac {1}{4\pi}B_iB_j\right\}\left(\frac{1}{\alpha}\frac{\partial }{\partial t}+{\bf V}\cdot\nabla \right )V^j\nonumber\\
+\rho_o\gamma^2V_i\left(\frac{1}{\alpha}\frac{\partial }{\partial t}+{\bf V}\cdot\nabla\right)\mu-\left(\frac{{\bf B}^2}{4\pi}\delta_{ij}-\frac {1}{4\pi}B_iB_j\right){V^j}_{,k}V^k \nonumber\\
=-\rho_o\mu\gamma^2a_i-p_{,i}+\frac {1}{4\pi}({\bf V}\times{\bf B})_i\nabla \cdot({\bf V}\times{\bf B})-\frac{1}{8\pi\alpha ^2}(\alpha{\bf B})^2_{,i}\nonumber\\
+\frac{1}{4\pi\alpha}(\alpha B_i)_{,j}B^j-\frac{1}{4\pi \alpha}[{\bf B}\times\{{\bf V}\times(\nabla \times(\alpha{\bf V}\times{\bf B}))\}]_i\label{eq17},
\end{eqnarray}
where a subscript $i$ on a vector quantity refers to the $i$-component of that vector. The $\mu\equiv(\rho+p)/\rho_o$ is the specific enthalpy of the fluid, where $\rho$ is the total density of mass-energy and $p$, the pressure as seen in the fluid's rest frame. The $\rho_o$ is the fluid's rest-mass density and $\gamma\equiv(1-\textbf{V}^2)^{-1/2}$ is the fluid's Lorentz factor as seen by the FIDOs. Equations (\ref{eq15})-(\ref{eq17}) are the perfect GRMHD equations for the RN black hole.

We assume for simplicity that the plasma has vanishing thermal pressure and vanishing thermal energy density; that is, we restrict our investigation to a \lq\lq cold plasma,\rq \rq for which we consider that the total density of mass-energy $\rho$ remains the same as the rest-mass density $\rho_o$:
\begin{equation}
p=0,\qquad\rho=\rho_o,\qquad \mbox{and} \qquad\mu=\frac{p+\rho}{\rho_o}=1\label{eq18}.
\end{equation}
Now using (\ref{eq18}) in (\ref{eq15})-(\ref{eq17}) we obtain the perfect GRMHD equations for cold plasma near to the event horizon of RN black hole.

We characterize the perturbed flow in the magnetosphere by its velocity $\bf V$ and magnetic field $\bf B$ as measured by the FIDOs, and the fluid's density $\rho$. The first order perturbations in these quantities are denoted by $\delta{\bf V}$, $\delta{\bf B}$ and $\delta\rho$. Accordingly, the perturbed variables take the following form:
\begin{equation}
\textbf{B}=\textbf{B}^o+\delta\textbf{B}, \qquad\textbf{V}=\textbf{V}^o+\delta\textbf{V}, \qquad\rho=\rho^o+\delta\rho\label{eq19},
\end{equation}
where $\textbf{B}^o$, $\textbf{V}^o$ and $\rho^o$ are unperturbed quantities. The waves can propagate in $z$-direction due to gravitation with respect to time $t$ and thus perturbed quantities must depend on $z$ and $t$.

\section{ Cold Plasma in Rotating Magnetized Background}\label{sec3}

We use the linear perturbation and Fourier analyze techniques to reduce GRMHD equations to ordinary differential equations. The magnetosphere has the perturbed flow along $x$-$z$ plane in this background. The FIDO-measured fluid four-velocity can be described in this plane by
\begin{equation}
\textbf{V}=V(z)\textbf{e}_{\textbf{x}} +u(z)\textbf{e}_{\textbf{z}}\label{eq20},
\end{equation}
while the Lorentz factor $\gamma$ takes the form
\begin{equation}
\gamma=\frac{1}{\sqrt{1-u^2-V^2}}\label{eq21}.
\end{equation}
The rotating magnetic field can be expressed in the $x$-$z$ plane as
\begin{equation}
{\bf B}= B[\lambda(z)\textbf{e}_{\textbf{x}} +\textbf{e}_{\textbf{z}}]\label{eq22}.
\end{equation}
The variables $\lambda$ , $u$ and $V$ are related by
\begin{equation}
V= \frac{V_F}{\alpha}+\lambda u\label{eq23},
\end{equation}
where $V_F$ is an integration constant.

We use the following perturbation quantities
\begin{equation}
{\tilde\rho}\equiv\delta \rho/\rho={\tilde\rho}(t,z),\;
{\bf v}\equiv\delta{\bf V}=v_x(t,z){\bf e}_x + v_z(t,z){\bf e}_z,\;{\bf b}\equiv \frac {\delta {\bf B}}{B}=b_x(t,z){\bf e}_x+b_z(t,z){\bf e}_z\label{eq24},
\end{equation}
with
\begin{eqnarray}
&&\tilde{\rho}(t,z)=c_1e^{-i(\omega t-kz)},\quad
v_z(t,z)=c_2e^{-i(\omega t-kz)},\quad
v_x(t,z)=c_3e^{-i(\omega t-kz)}\nonumber\\
&&b_x(t,z)=c_4e^{-i(\omega t-kz)},\quad b_z(t,z)=c_5e^{-i(\omega t-kz)}
\label{eq25},
\end{eqnarray}
where $c_s$, $s=1,\cdots,5$, are arbitrary constants.

Using (\ref{eq24}) in (\ref{eq15})-(\ref{eq17}), we obtain
\begin{equation}
\frac{\partial(\delta {\bf B})}{\partial t}=\nabla\times (\alpha\textbf{v}\times\textbf{B})+\nabla\times(\alpha\textbf{V} \times\delta\textbf{B})\label{eq26},
\end{equation}
\begin{equation}
\nabla\cdot(\delta\textbf{B})=0\label{eq27},
\end{equation}
\begin{eqnarray}
\left(\frac{1}{\alpha}\frac {\partial}{\partial t} + {\bf V}\cdot \nabla \right )\delta \rho+ \rho \gamma ^2 {\bf V}\cdot \left ( \frac {1}{\alpha }\frac {\partial}{\partial t} + {\bf V}\cdot \nabla \right ){\bf v}-\frac {\delta \rho }{\rho}({\bf V}\cdot \nabla)\rho + \rho (\nabla \cdot {\bf v})\nonumber\\
=-2 \rho \gamma ^2({\bf V \cdot v})({\bf V}\cdot \nabla){\ln} \gamma - \rho \gamma ^2 ({\bf V}\cdot \nabla{\bf V})\cdot{\bf v}+ \rho ({\bf v} \cdot \nabla\ln u)\label{eq28},
\end{eqnarray}
\begin{eqnarray}
&&\left\{\left(\rho\gamma ^2 +\frac {{\bf B}^2}{4\pi}\right )\delta _{ij}+\rho \gamma ^4V_iV_j-\frac {1}{4\pi}B_iB_j\right\}\frac {1}{\alpha}\frac {\partial v^j}{\partial t}+ \frac {1}{4\pi}\left [{\bf B}\times \left\{{\bf V}\times \frac {1}{\alpha}\frac {\partial (\delta {\bf B})}{\partial t}\right\}\right]_i \nonumber\\
&&+ \rho \gamma ^2v_{i,j}V^j+  \rho \gamma ^4v_{j,k}V_iV^jV^k - \frac {1}{4\pi \alpha}\{(\alpha \delta B_i)_{,j} - (\alpha \delta B_j)_{,i} \}B^j \nonumber\\
&&= -\gamma^2 \{\delta \rho + 2 \rho \gamma ^2({\bf V \cdot v})\}a_i + \frac {1}{4\pi \alpha }\{(\alpha B_i)_{,j} -(\alpha B_j)_{,i} \}\delta B^j \nonumber\\
&&- \rho \gamma ^4 (v_iV^j + v^jV_i)V_{k,j}V^k - \gamma ^2 \{\delta \rho V^j + 2\rho \gamma^2({\bf V \cdot v})V^j + \rho v^j \}V_{i,j} \nonumber\\
&&-\gamma ^4V_i \{\delta \rho V^j + 4\rho \gamma ^2({\bf V \cdot v})V^j + \rho v^j \}V_{j,k}V^k\label{eq29}.
\end{eqnarray}
The component form of (\ref{eq26})-(\ref{eq29}) can be written as follows :
\begin{eqnarray}
\frac {1}{\alpha }\frac {\partial {b_x}}{\partial t}+ub_{x,z}&=&\nabla\ln\alpha (v_x-{\lambda }v_z +Vb_z-ub_x)\nonumber\\
&&+(v_{x,z}-{\lambda }v_{z,z}-\lambda 'v_z +V'b_z-u'b_x)\label{eq30},\\
\frac{1}{\alpha}\frac{\partial b_z}{\partial t}+ub_{z,z}&=&0\label {eq31},\\
b_{z,z}&=&0\label{eq32},
\end{eqnarray}
\begin{eqnarray}
&&\frac {1}{\alpha}\frac {\partial {\tilde \rho}}{\partial t} + u {\tilde \rho}_{,z} + \gamma^2 V \left (\frac {1}{\alpha}\frac {\partial v_x}{\partial t}+ uv_{x,z}\right )+ \gamma^2 u\frac {1}{\alpha}\frac {\partial v_z}{\partial t} + (1+\gamma^2u^2)v_{z,z}\nonumber\\
&&=-\gamma^2 u\{(1+2\gamma^2V^2)V'+ 2 \gamma^2uVu'\}v_x \nonumber\\
&&+\left\{(1-2\gamma^2u^2)(1+\gamma^2u^2)\frac {u'}{u}- 2\gamma^4u^2VV'\right\}v_z\label{eq33},
\end{eqnarray}
\begin{eqnarray}
&&\left\{\rho \gamma ^2 (1+\gamma ^2V^2)+\frac {\textbf{B}^2}{4\pi }\right\}\frac {1}{\alpha }\frac {\partial v_x}{\partial t}+ \left ( \rho \gamma ^4uV-\frac {\lambda\textbf{B}^2}{4\pi }\right )\frac {1}{\alpha }\frac {\partial v_z}{\partial t}\nonumber\\
&&+\left\{\rho \gamma ^2 (1+\gamma ^2V^2)-\frac {\textbf{B}^2}{4\pi }\right\} uv_{x,z} + \left (\rho \gamma ^4uV + \frac {\lambda\textbf{B}^2}{4\pi }\right )uv_{z,z}- \frac {\textbf{B}^2}{4\pi }(1-u^2)b_{x,z}\nonumber\\
&&-\frac {\textbf{B}^2}{4\pi \alpha }b_x \{{\alpha '}(1-u^2)- \alpha uu'\}+ {\tilde \rho}\rho \gamma ^2u\{(1+\gamma ^2V^2)V'+ \gamma ^2uu'V \}\nonumber\\
&&+\left[\rho \gamma ^4u\{(1+4\gamma^2V^2)uu' + 4(1+\gamma^2V^2)VV'\}+ \frac {\textbf{B}^2u\alpha'}{4\pi \alpha}\right]v_x\nonumber\\
&&+\left[\rho\gamma^2\left[\{(1+2\gamma^2u^2) (1+2\gamma^2V^2)-\gamma^2V^2\}
V'\right.\right.\nonumber\\
&&\left.\left.+ 2\gamma^2 (1+2\gamma^2u^2)uu'V\right]+\frac{\textbf{B}^2u}{4\pi \alpha}(\alpha \lambda)'\right]v_z=0\label{eq34},
\end{eqnarray}
\begin{eqnarray}
&&\left\{\rho \gamma ^2 (1+\gamma ^2u^2)+\frac {\lambda ^2 \textbf{B}^2}{4\pi }\right\}\frac {1}{\alpha }\frac {\partial v_z}{\partial t}+ \left ( \rho \gamma ^4uV-\frac {\lambda\textbf{B}^2}{4\pi }\right )\frac {1}{\alpha }\frac {\partial v_x}{\partial t}\nonumber\\
&&+\left\{\rho \gamma ^2 (1+\gamma ^2u^2)-\frac {\lambda ^2 \textbf{B}^2}{4\pi }\right\} uv_{z,z} + \left (\rho \gamma ^4uV + \frac {\lambda\textbf{B}^2}{4\pi }uv_{x,z}\right )\nonumber\\
&&+ \frac {\lambda\textbf{B}^2}{4\pi }(1-u^2)b_{x,z}+\frac{\textbf{B}^2}{4\pi\alpha}\{(\alpha \lambda)^\prime+ \alpha^\prime\lambda -u\lambda (u\alpha^\prime+u^\prime\alpha)\}b_x \nonumber\\
&&+{\tilde\rho}\gamma^2[a_z+u\{(1+\gamma ^2u^2)u^\prime+ \gamma ^2uVV^\prime\}]\nonumber\\
&&+\left[\rho \gamma ^4\{u^2V^\prime(1+4\gamma^2V^2)+2V(a_z+uu^\prime(1+2\gamma^2u^2))\}+ \frac {\lambda\textbf{B}^2u\alpha^\prime}{4\pi \alpha}\right]v_x\nonumber\\
&&+[\rho\gamma^2\{u^\prime(1+\gamma^2u^2)(1+4\gamma^2u^2)\nonumber\\
&&+2u\gamma^2\{(1+2\gamma^2u^2)VV^\prime+a_z\}\}- \frac {\lambda\textbf{B}^2u }{4\pi \alpha}(\alpha \lambda)^\prime]v_z=0\label{eq35}.
\end{eqnarray}
From the Fourier analyzed of (\ref{eq30})-(\ref{eq35}) with (\ref{eq25}), we obtain
\begin{equation}
c_3(\alpha^\prime+ik\alpha)-c_2\{(\alpha\lambda)^\prime+ik\alpha \lambda\}+c_5(\alpha V)^\prime-c_4\{(\alpha u)^\prime-i\omega +ik\alpha u\}=0\label{eq36},
\end{equation}
\begin{equation}
c_5 \left (-\frac {i\omega }{\alpha } +iku \right )=0\label{eq37},
\end{equation}
\begin{equation}
c_5 ik=0\label{eq38},
\end{equation}
\begin{eqnarray}
&&+c_1\left(-\frac {i\omega }{\alpha }+iku \right)+c_2\left\{-\frac{i\omega}{\alpha}\gamma^2u+ik(1+\gamma ^2u^2)\right.\nonumber\\
&&\left.-(1-2\gamma^2u^2)(1+\gamma^2u^2)\frac{u^\prime}{u}+2\gamma ^4u^2VV^\prime\right\}\nonumber\\
&&+c_3\gamma^2\left[\left(-\frac{i\omega}{\alpha}+iku\right)V + u\{(1+2\gamma^2V^2)V^\prime+2\gamma^2uu^\prime V\}\right]=0 \label{eq39}
\end{eqnarray}
\begin{eqnarray}
&&c_1\rho\gamma^2u\{(1+\gamma^2V^2)V^\prime+\gamma^2uu^\prime V\}- c_4\frac{\textbf{B}^2}{4\pi}\left\{(1-u^2)ik +\frac{\alpha^\prime}{\alpha}(1-u^2)-uu^\prime\right\}\nonumber\\
&&+c_2\left[-\left(\rho\gamma^4uV-\frac{\lambda\textbf{B}^2}{4\pi }\right)\frac{i\omega}{\alpha}+iku\left(\rho\gamma^4uV+\frac {\lambda\textbf{B}^2}{4\pi}\right)\right.\nonumber\\
&&\left.+\rho\gamma^2\{(1+2\gamma ^2u^2)(1+2\gamma ^2V^2)-\gamma ^2V^2\}V'\right.\nonumber\\
&&\left.+2\rho\gamma^4(1+2\gamma^2u^2)uu^\prime V+\frac {\textbf{B}^2u}{4\pi\alpha}(\lambda\alpha)^\prime\right]+c_3 \left[-\left\{\rho\gamma^2(1+\gamma^2V^2)+\frac{\textbf{B}^2}{4\pi }\right\}\frac{i\omega}{\alpha}\right.\nonumber\\
&&+iku\left\{\rho\gamma^2(1+\gamma^2V^2)-\frac{\textbf{B}^2}{4\pi }\right\}\nonumber\\
&&\left.+\rho\gamma^4u\{(1+4\gamma^2V^2)uu^\prime+4(1+\gamma ^2V^2)VV^\prime\}-\frac{\textbf{B}^2u\alpha^\prime}{4\pi\alpha }\right]=0\label{eq40},
\end{eqnarray}
\begin{eqnarray}
&&c_1\rho\gamma^2[a_z+u\{(1+\gamma ^2u^2)u^\prime+\gamma ^2uVV^\prime\}]\nonumber\\
&&+c_2\left[-\left\{\rho\gamma^2(1+\gamma^2u^2)+\frac{\lambda ^2\textbf{B}^2}{4\pi}\right\}\frac{i\omega}{\alpha }+\left\{\rho\gamma^2(1+\gamma ^2u^2)-\frac{\lambda^2\textbf{B}^2}{4\pi }\right\}iku\right.\nonumber\\
&&\left.+\left\{\rho\gamma^2\{u^\prime(1+\gamma ^2u^2)(1+4\gamma ^2u^2)+2u\gamma^2((1+2\gamma^2u^2)VV'+a_z)\}-\frac{\lambda  \textbf{B}^2u}{4\pi\alpha}(\alpha\lambda )^\prime\right\}\right]\nonumber\\
&&c_3\left[-\left(\rho\gamma^4uV-\frac {\lambda\textbf{B}^2}{4\pi }\right )\frac {i\omega }{\alpha } +iku \left(\rho \gamma^4uV+\frac{\lambda\textbf{B}^2}{4\pi }\right )\right.\nonumber\\
&&\left.+\left\{\rho \gamma ^4\{u^2V^\prime(1+4\gamma^2V^2)+ 2V(a_z+uu^\prime(1+2\gamma ^2u^2))\}+\frac{\lambda  \textbf{B}^2u\alpha^\prime}{4\pi \alpha }\right\}\right]\nonumber\\
&&+\frac{\textbf{B}^2}{4\pi}c_4\left[\lambda (1-u^2)ik+\lambda (1-u^2)\frac{\alpha^\prime}{\alpha}-\lambda uu^\prime+ \frac{(\lambda\alpha)^\prime}{\alpha}\right]=0.\label{eq41}
\end{eqnarray}
Equation (\ref{eq38}) implies that $c_5$ is zero which gives that $b_z$ is zero. The determinant of the coefficients of $c_1$, $c_2$, $c_3$ and $c_4$ in (\ref{eq36}) and (\ref{eq39})-(\ref{eq41}) equated to zero gives a complex relation in $k$, called the dispersion relation, which is of the form $A(z,\omega)k^4+B(z,\omega)k^3+C(z,\omega)k^2+D(z,\omega)k +E(z,\omega)=0$.\\

\noindent
\textbf{Numerical Solution Modes}\\
We investigate the different types of modes of waves when $\textbf{B}>0$ and the wave number is in arbitrary direction to $\textbf{B}$. We use the lapse function $\alpha=\frac{z}{2r_+} \left(= \frac{z}{2(M+\sqrt{M^2-Q^2})}\right)$, $0\le Q^2/M^2\le 1$. We consider the black hole mass $M\sim1M_\odot$, $Q^2/M^2=0.7$, $\rho =1$ and $\frac {\textbf{B}^2}{4\pi}=2$. From the mass conservation law in three-dimensions we get $u=\frac{1}{\sqrt{2+z^2}}$. For simplicity, we also assume that $u=V$. From (\ref{eq23}) we get $ \lambda  =1- \frac {\sqrt {2+z^2}}{z} $ by taking $V_F = 1$, which shows that the magnetic field diverges close to the horizon.

Using these values in the dispersion relation we get values for $k$, from which we have calculated the phase velocity $v_p\equiv\frac{\omega}{k}$ and group velocity $v_g\equiv(n+\omega\frac{dn}{d\omega})^{-1}$, where $n(=1/v_p)$ is the refractive index computed as the ratio of the speed of light in a vacuum to the speed of light through the material and $\frac {dn}{d\omega}$ determines whether the dispersion is normal or not.

The real part of the dispersion relation gives four real values for $k$ , out of which two are real and interesting. The other two values turn out to be imaginary in the whole region. One real value for $k$ is obtained out of three values found from imaginary part of dispersion relation. This is shown by Fig. 3. The two dispersion relations obtained from the real part are shown in the Fig. 1 and Fig. 2.

\begin{figure}[h]
\begin{center}
 \includegraphics[scale=.6]{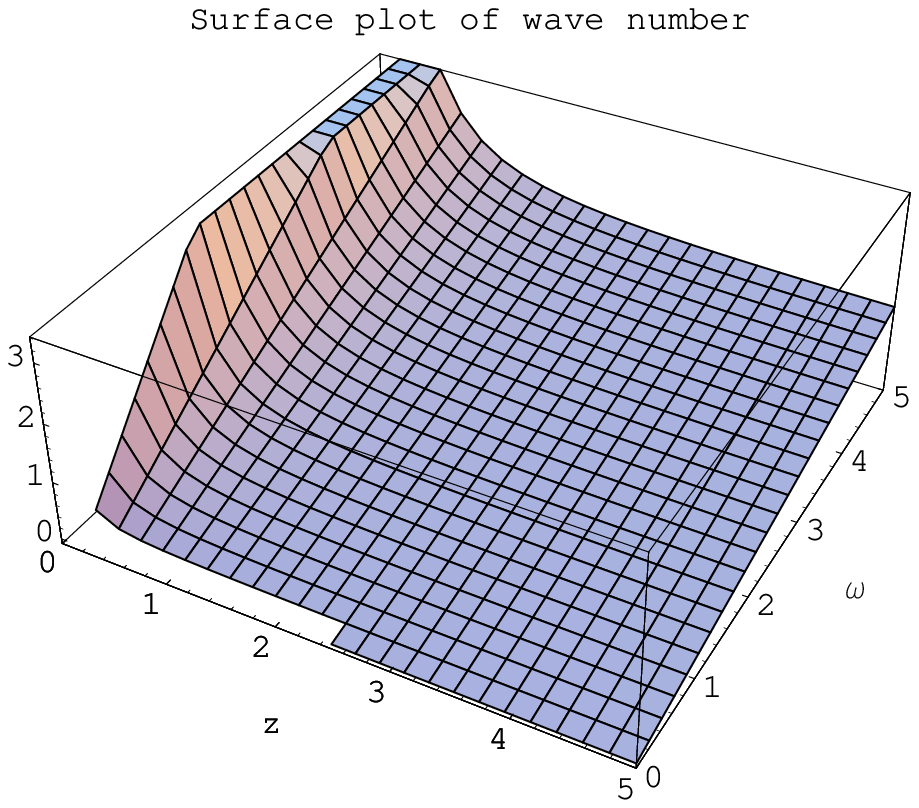}\\
 \includegraphics[scale=.6]{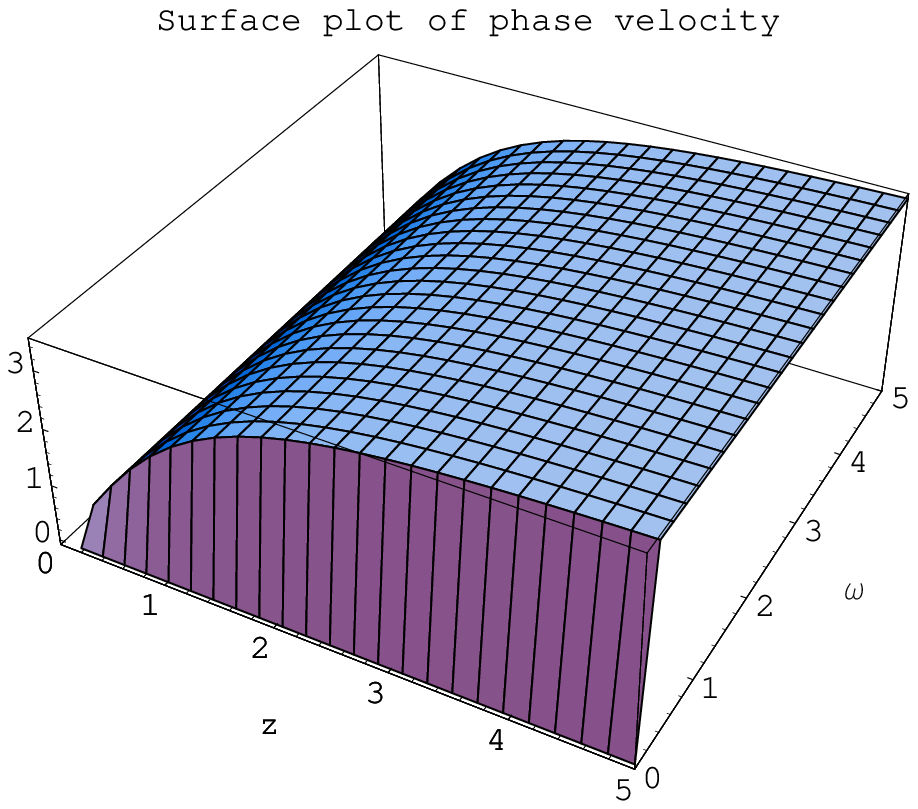}
 \includegraphics[scale=.6]{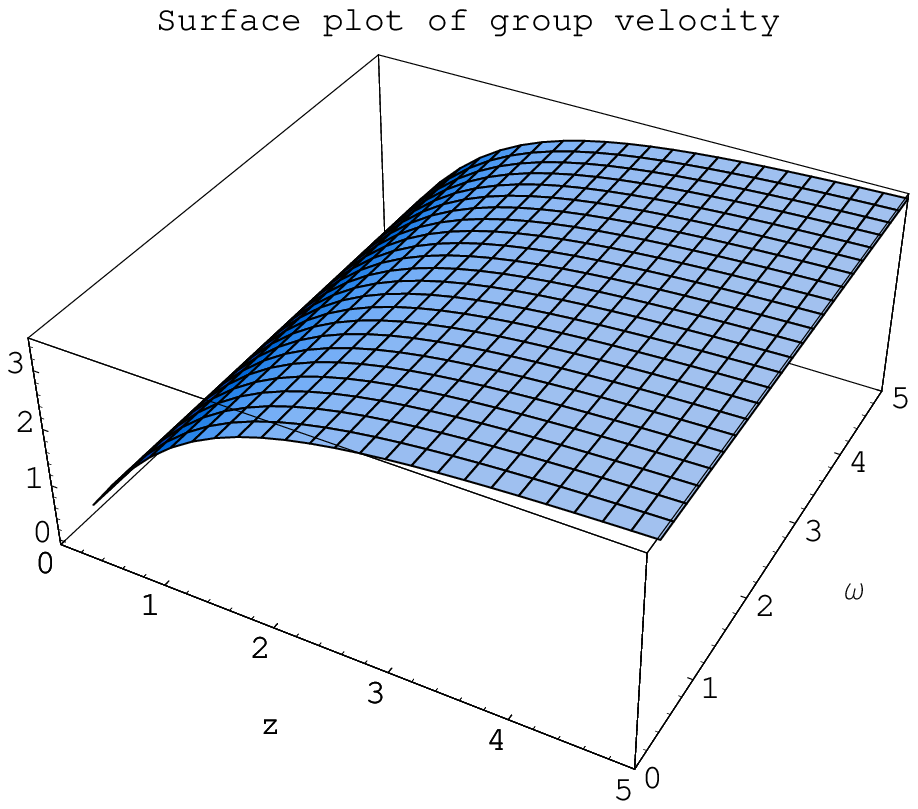}\\
 \includegraphics[scale=.6]{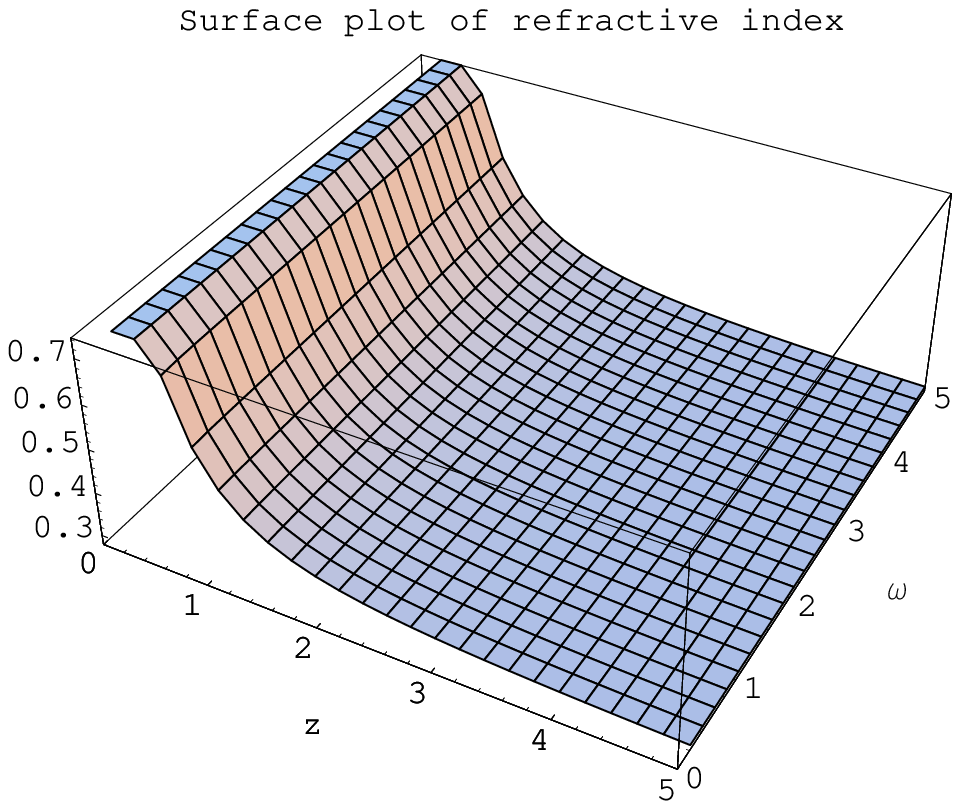}
 \includegraphics[scale=.6]{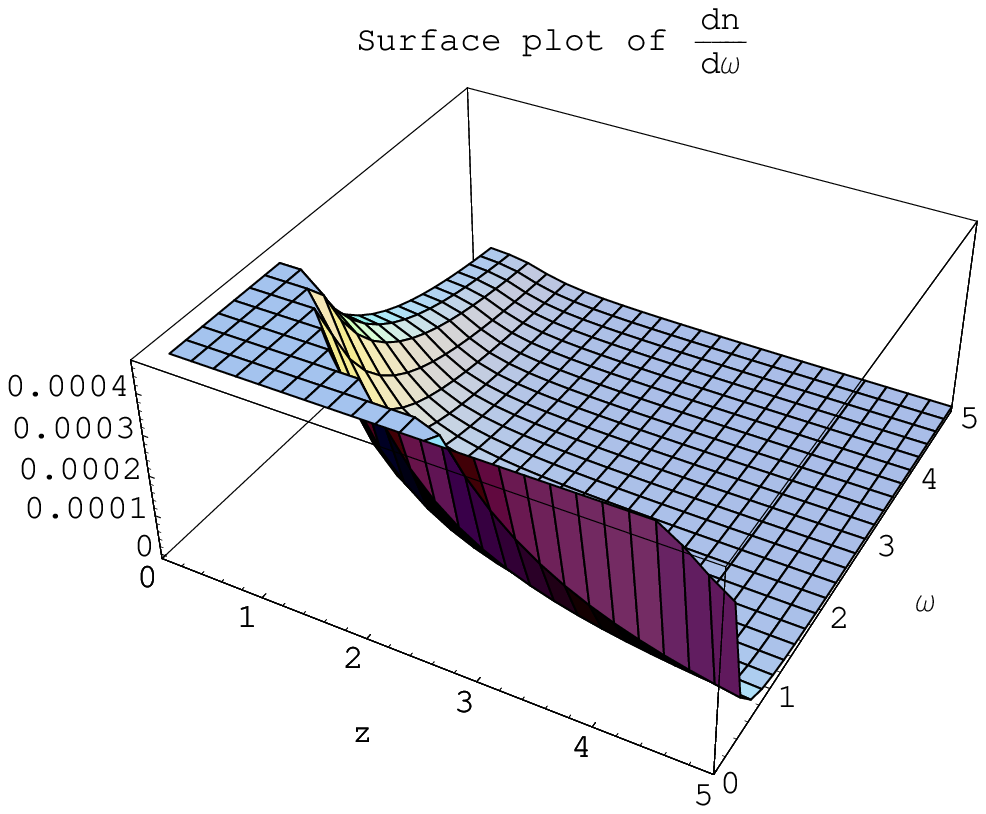}
\end{center}
\caption{The region is of non-normal dispersion, phase and group velocities are the same and increasing with increase of $z$,  $\frac {dn}{d\omega }>0$ but $n<1$.}
\end{figure}

We see from Fig. 1 that the waves gain energy with the increase in angular frequency but lose when we move away from the horizon and hence damping arises. But in the vicinity of the horizon the wave number is very large due to strong gravitational field, so no wave exists there. The phase and group velocities increase as we depart from the horizon.  The dispersion is not normal there because of refractive index $n<1$ though $\frac {dn}{d\omega}>0$.

\begin{figure}[h]
\begin{center}
 \includegraphics[scale=.6]{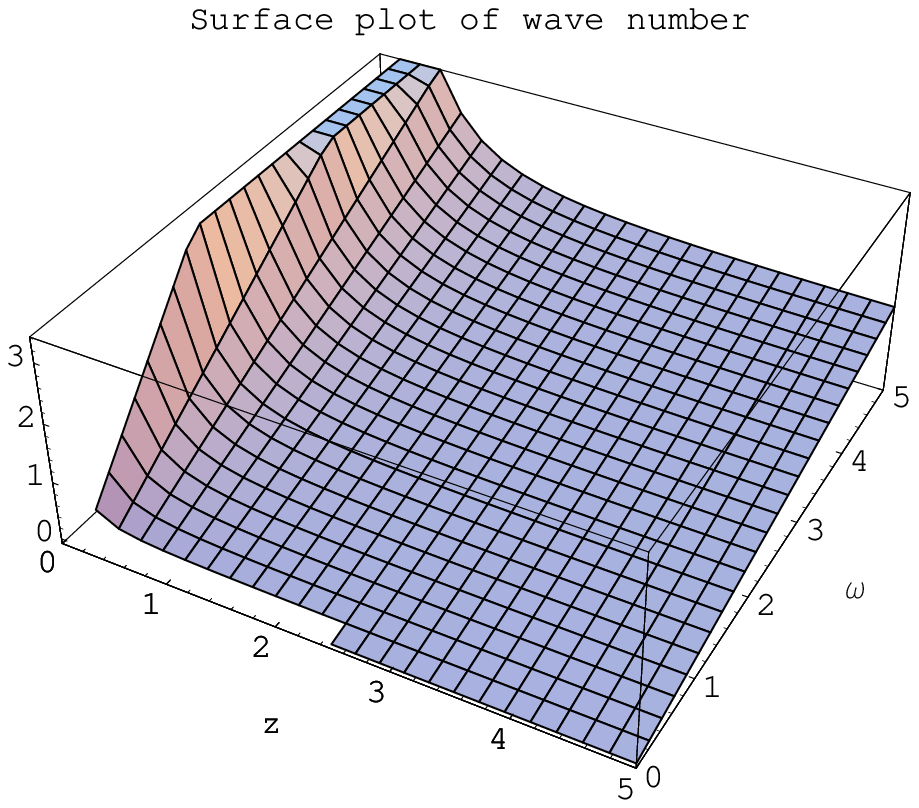}\\
 \includegraphics[scale=.6]{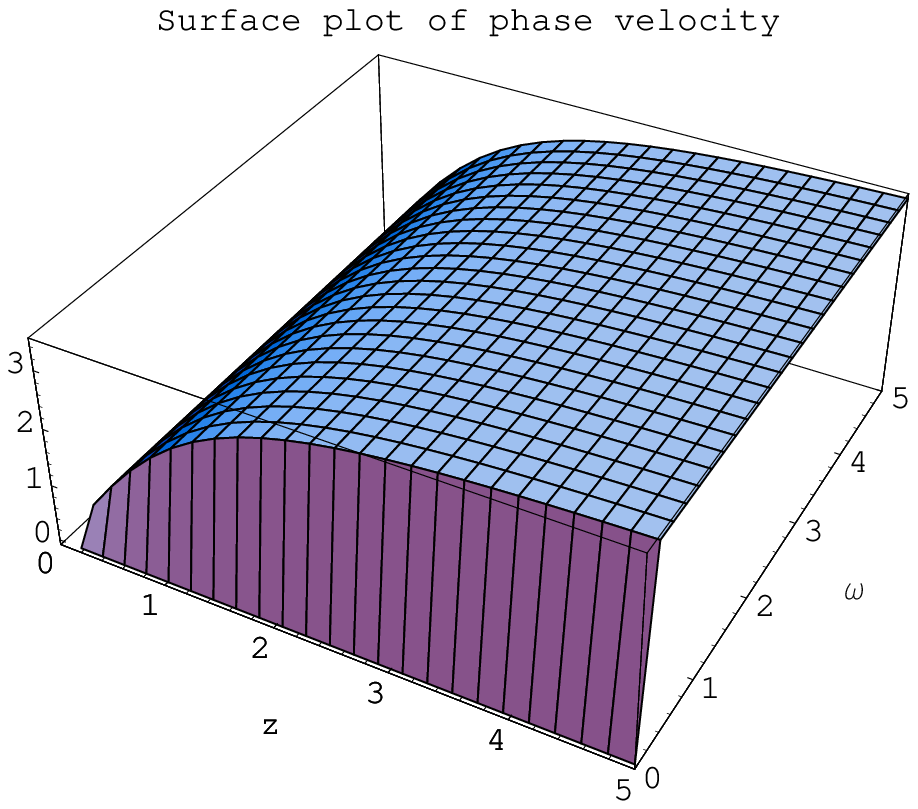}
 \includegraphics[scale=.6]{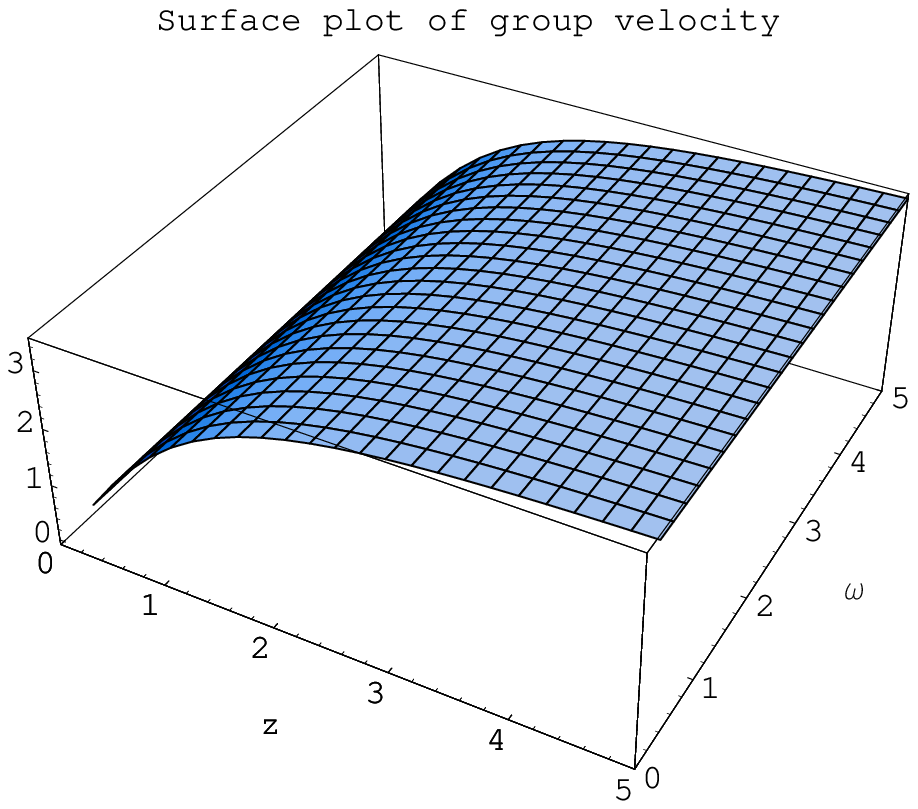}\\
 \includegraphics[scale=.6]{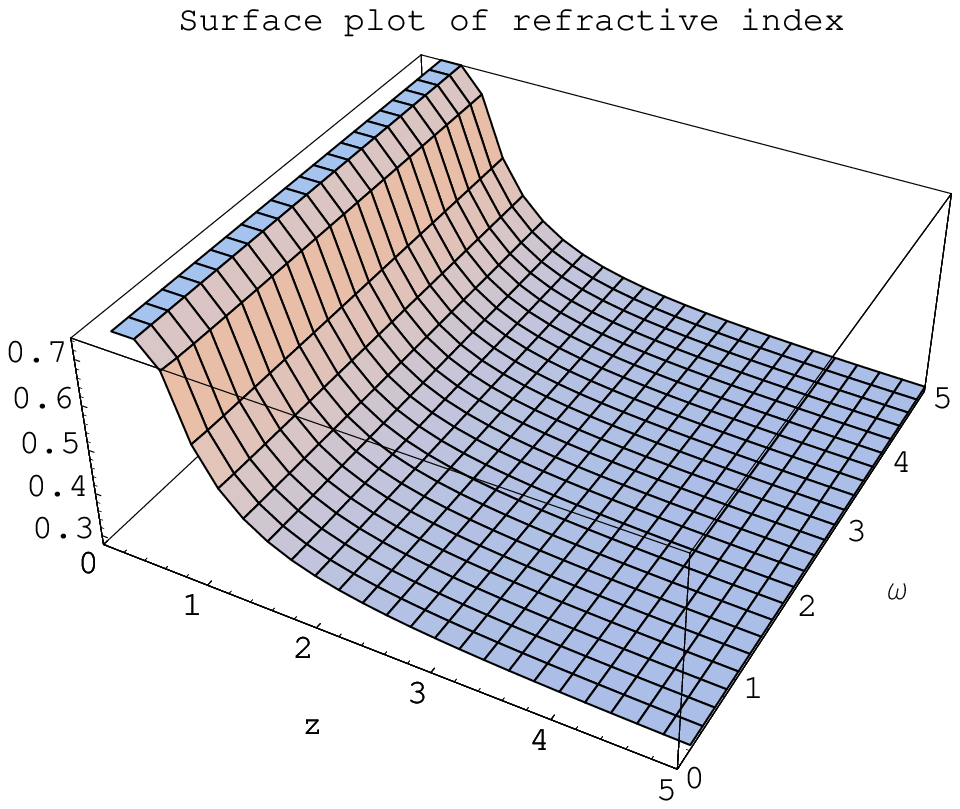}
 \includegraphics[scale=.6]{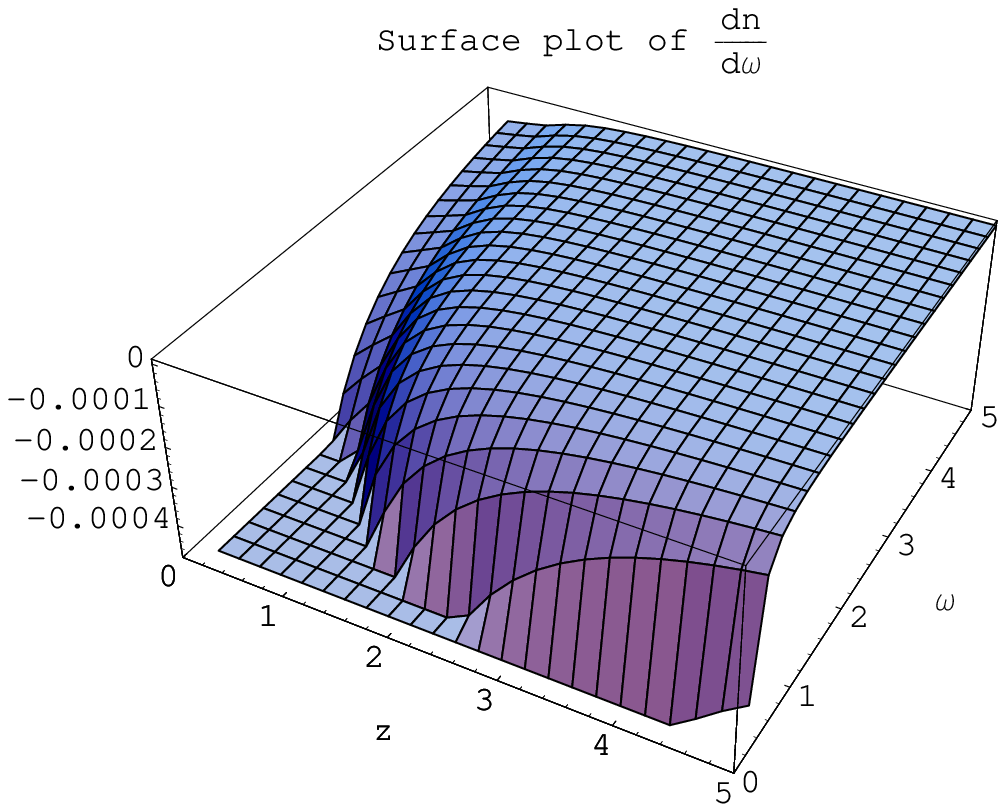}
\end{center}
\caption{The region is of non-normal dispersion, phase and group velocities are the same, $n<1$ and $\frac{dn}{d\omega}<0$.}
\end{figure}

The Fig. 2 indicates that the wave number is proportional to angular frequency and inversely proportional to $z$. There exists no wave at the horizon since wave number is infinite there. The phase and group velocities show the same behavior. We observe the damping modes as we depart from the event horizon and the growing modes as $z$ decreases. Since $n<1$ and $\frac {dn}{d\omega }<0$, the wave does not disperse normally there.

\begin{figure}[h]
\begin{center}
 \includegraphics[scale=.6]{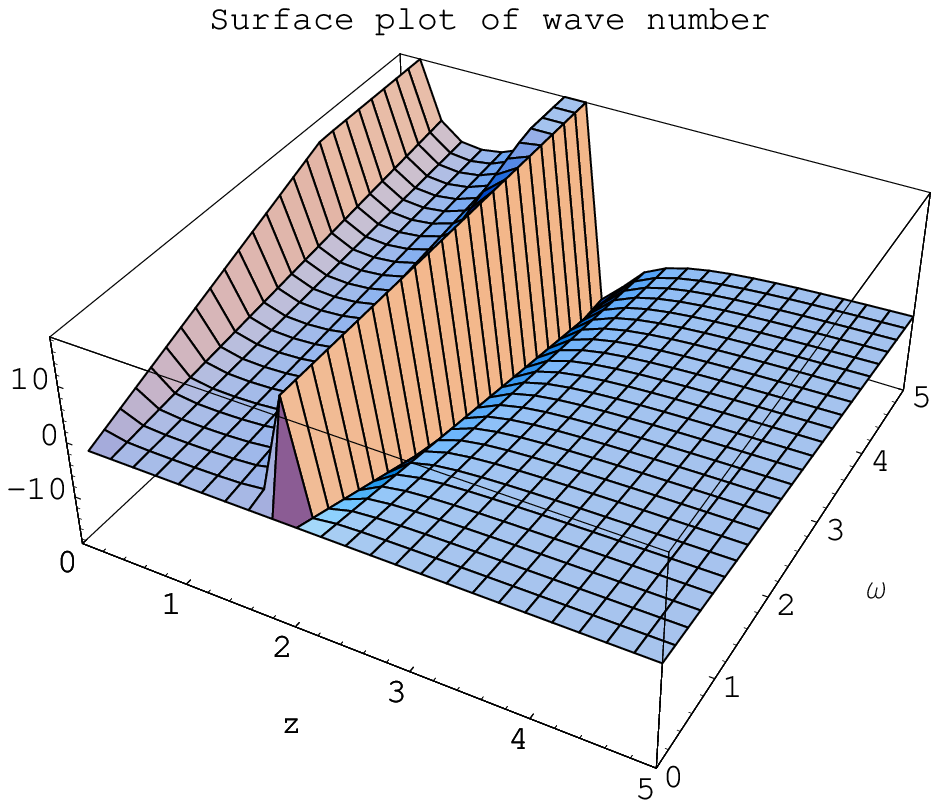}\\
 \includegraphics[scale=.6]{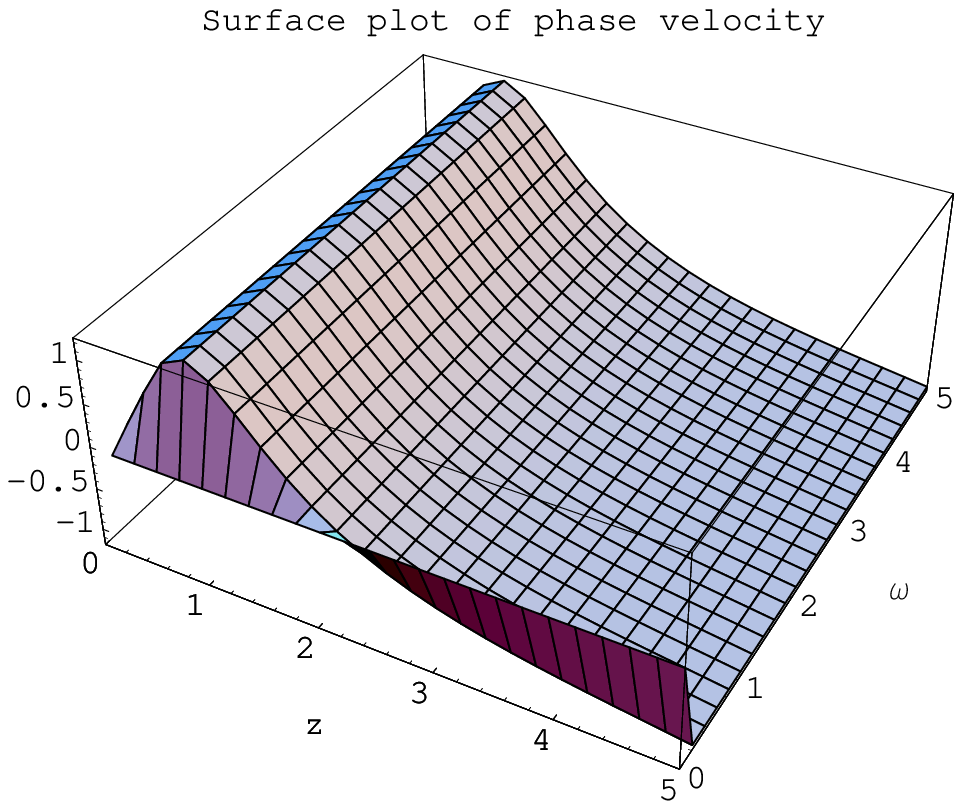}
 \includegraphics[scale=.6]{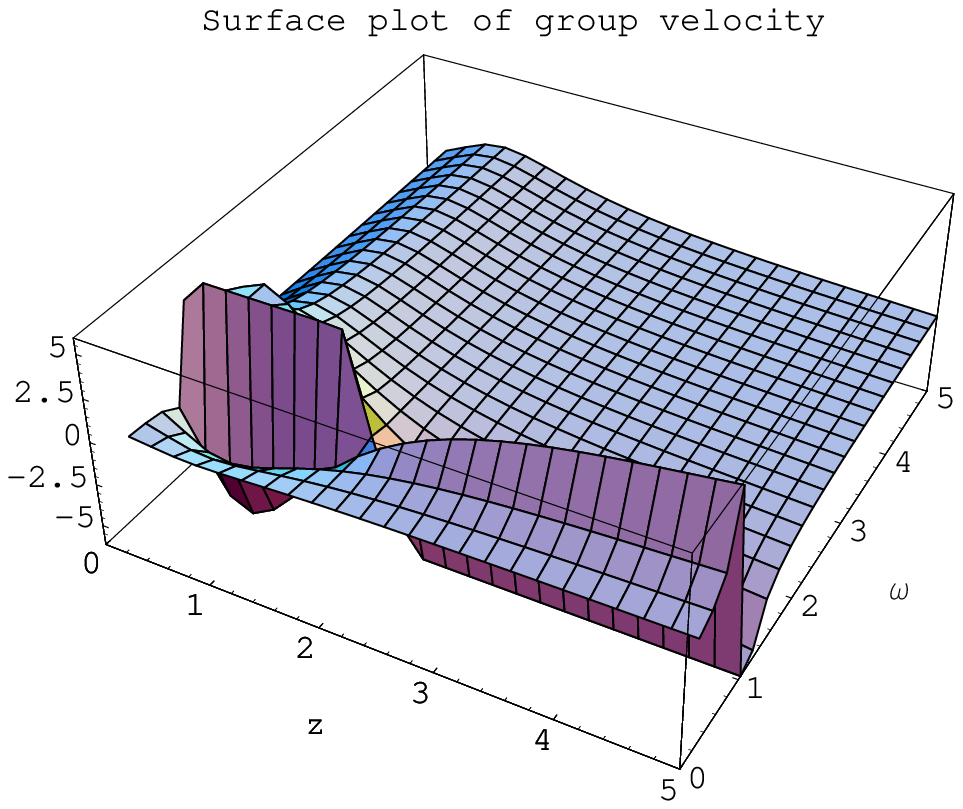}\\
 \includegraphics[scale=.6]{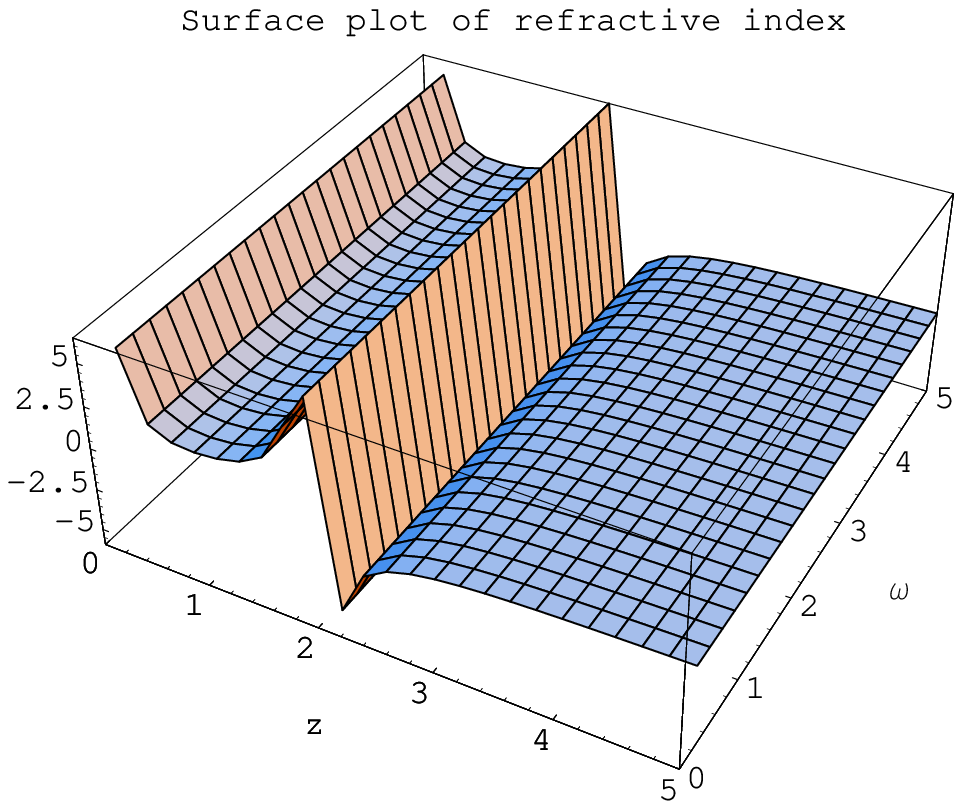}
 \includegraphics[scale=.6]{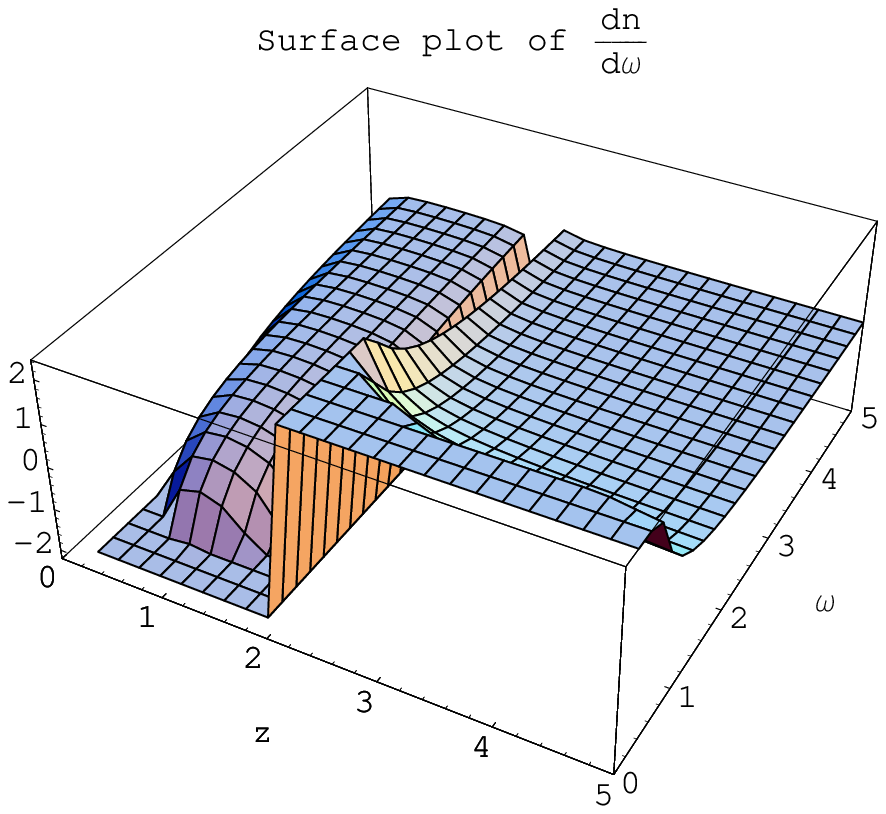}
\end{center}
\caption{The region is of anomalous dispersion, $v_g>v_p$, and $k,v_p,v_g$ are negative, $n<1, \frac{dn}{d\omega}<0$. The medium has the properties of metamaterials.}
\end{figure}

Figure 3 shows that the wave number is very large at the event horizon ($z=0$) but it decreases as we depart from the horizon and also it takes negative value for some values of $z$ and $\omega$. The phase and group velocities are also negative and the group velocity is greater than the phase velocity. The value of refractive index $n<1$ and $\frac {dn}{d\omega }<0$ for some values of $z$ and $\omega$. Therefore the region is of anomalous dispersion and possesses the properties of metamaterials.

\section{Cold Plasma in Rotating Non-magnetized Background}\label{sec4}

In non-magnetized background $\textbf{B}=\textbf{0}$ i.e. when $b_x=b_z=0$, the GRMHD equations (\ref{eq36})-(\ref{eq41}) reduce to the following form:
\begin{eqnarray}
&&c_1\left(-\frac{i\omega}{\alpha}+iku\right)+c_2\left[-\frac {i\omega}{\alpha}\gamma^2u+(1+\gamma^2u^2)ik- (1-2\gamma^2u^2)(1+\gamma^2u^2)\frac {u^\prime}{u}+ 2\gamma^4u^2VV^\prime\right]\nonumber\\
&&+c_3\gamma^2\left[-\frac{i\omega}{\alpha}V+ikuV +\gamma^2u\{(1+2\gamma^2V^2)V^\prime+2\gamma^2uu^\prime V\}\right]=0,\label{eq42}
\end{eqnarray}
\begin{eqnarray}
&&c_1\gamma^2u\{(1+\gamma^2V^2)V^\prime+\gamma^2uu^\prime V\}+c_2\gamma^2\left[\left(-\frac{i\omega}{\alpha}+ iku\right)\gamma^2uV\right.\nonumber\\
&&+\left.\{(1+2\gamma^2u^2)(1+2\gamma^2V^2)- \gamma^2V^2\}V^\prime+2\gamma^2 (1+2\gamma^2u^2)uu^\prime V\right]\nonumber\\
&&+c_3\left[\left(-\frac{i\omega}{\alpha}+iku\right)\gamma^2 (1+\gamma^2V^2)\right.\nonumber\\
&&\left.+\gamma^4u\left\{(1+4\gamma^2V^2)uu^\prime +4VV^\prime(1+\gamma^2V^2)\right\}\right]=0\label {eq43},
\end{eqnarray}
\begin{eqnarray}
&&c_1\gamma^2\{a_z+(1+\gamma^2u^2)uu^\prime+\gamma^2u^2VV^\prime\}+c_2 [\gamma^2(1+\gamma^2u^2)\left(-\frac{i\omega}{\alpha} +iku\right)\nonumber\\
&&+\gamma^2\{u^\prime(1+\gamma^2u^2)(1+4\gamma^2u^2)+2u\gamma^2(a_z +(1+2\gamma^2u^2)VV^\prime)\}]\nonumber\\
&&+c_3\gamma^4V\left[\left(-\frac{i\omega}{\alpha}+iku\right)u +u^2(1+4\gamma^2V^2)+2\{a_z +uu^\prime(1+2\gamma^2u^2)\}\right]=0\label{eq44}.
\end{eqnarray}
where the FIDO-measured fluid four-velocity $\textbf{V}$, Lorentz factor $\gamma$ are given by (\ref{eq20}) and (\ref{eq21}), respectively. The determinant of the coefficients $c_1$, $c_2$ and $c_3$ in (\ref{eq42})-(\ref{eq44}) yields a complex dispersion relation of the form $A(z,\omega)k^3+B(z,\omega)k^2+C(z,\omega)k+D(z,\omega)=0$.

To analyze the numerical solution mode we consider $Q^2/M^2=0.5$ and assume that $V=u$ and $\rho=$constant. The mass conservation law gives that $u=\frac{1}{\sqrt{z^2+2}}$. From the real part of the dispersion relation we get only two real values for $k$ which are shown in Fig. 4 and Fig. 5. The imaginary part gives only one real root and  two complex conjugate roots. Figure 6 shows the real value for $k$ obtained from the imaginary part of the dispersion equation.

\begin{figure}[h]
\begin{center}
 \includegraphics[scale=.6]{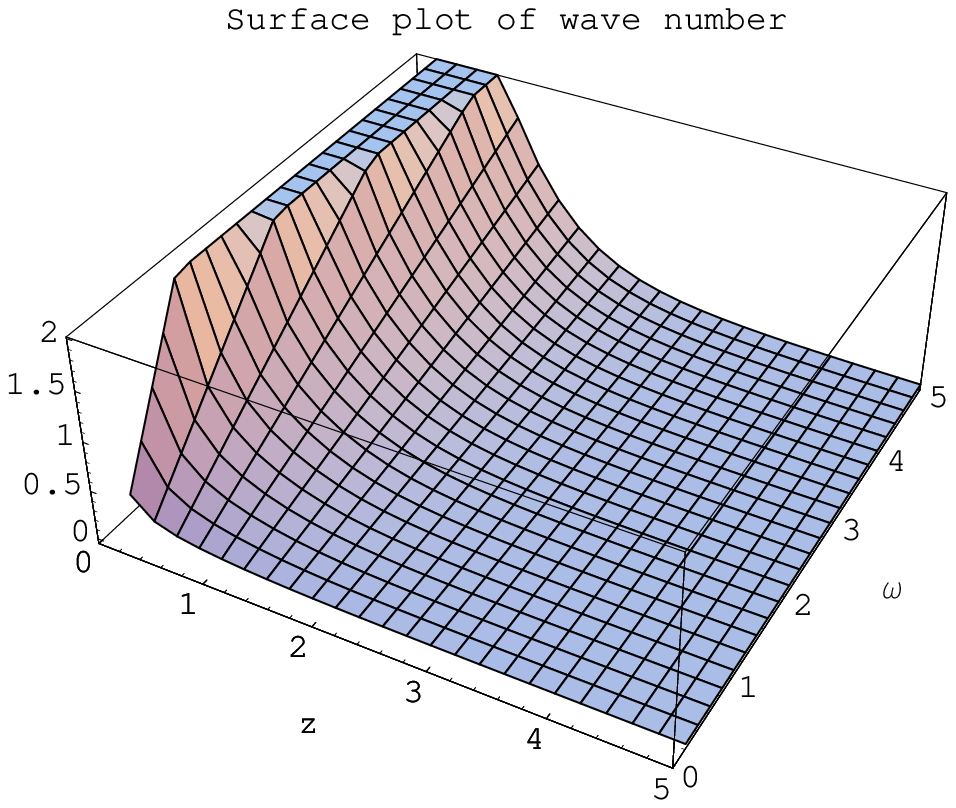}\\
 \includegraphics[scale=.6]{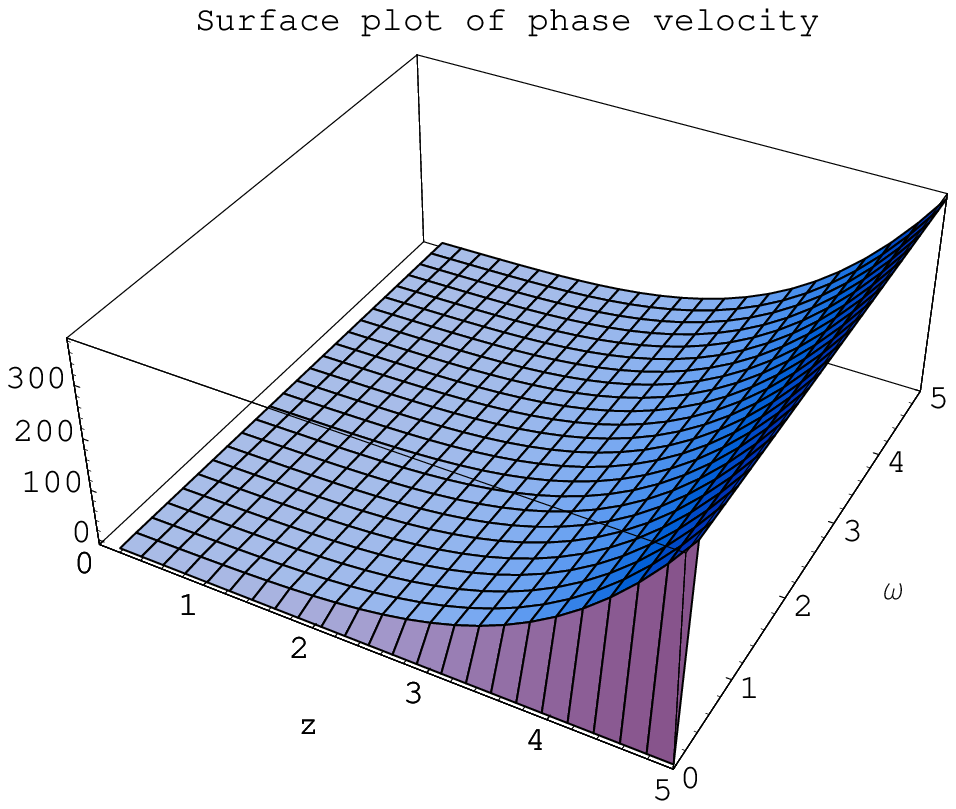}
 \includegraphics[scale=.6]{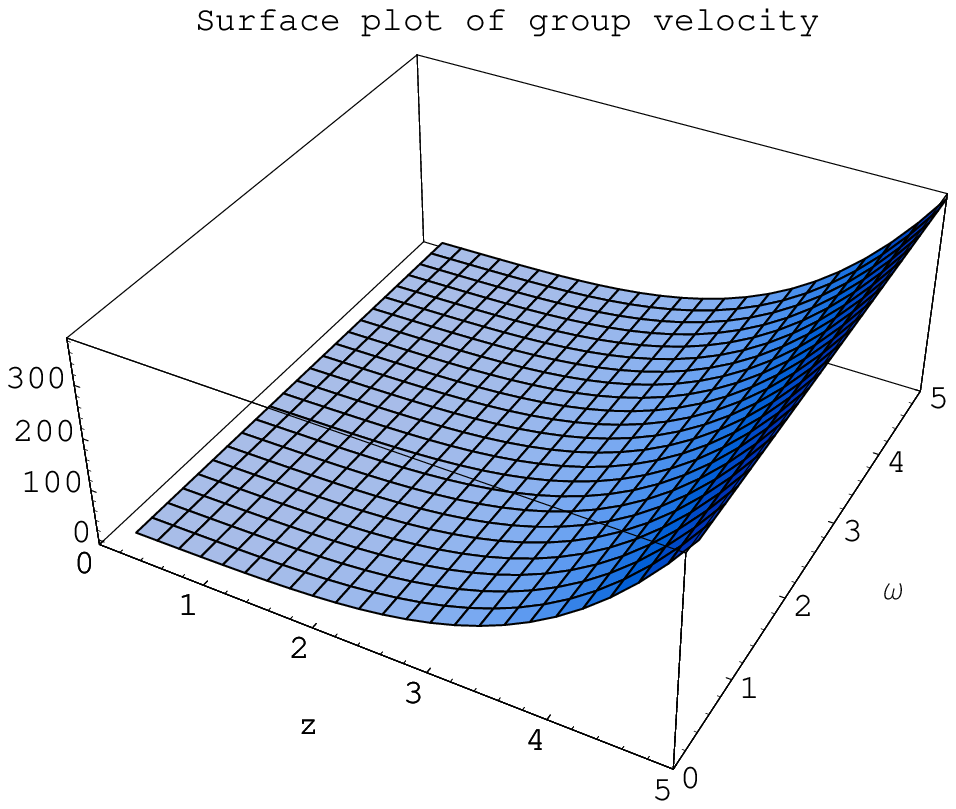}\\
 \includegraphics[scale=.6]{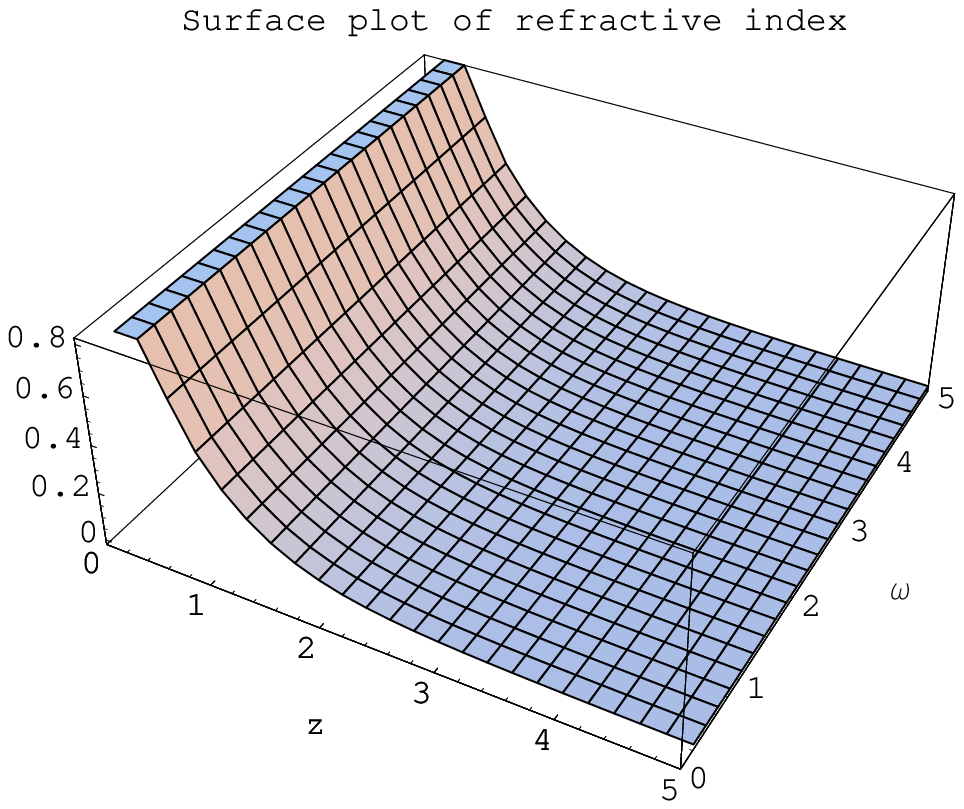}
 \includegraphics[scale=.6]{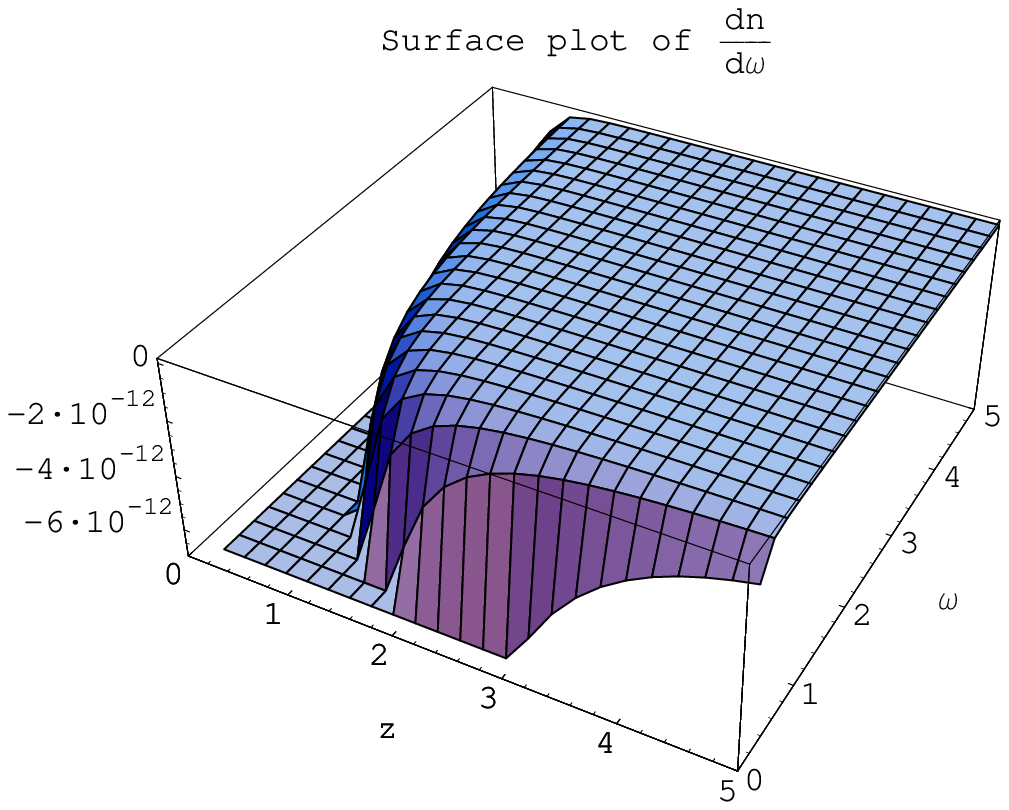}
\end{center}
\caption{The region is of non-normal dispersion, phase and group velocities are the same,  $n<1$ and $\frac {dn}{d\omega }<0$.}
\end{figure}

We see from Fig. 4 that the wave number is very large close to the event horizon and the waves lose energy as we move away from the horizon of RN black hole. This shows that the increase in $\omega$ increases $k$ and the waves are in growing mode as $z$ decreases. The phase and group velocities are of the same pattern and increase as $z$ increases. Since $n<1$ and $\frac{dn}{d\omega}<0$, the region is not of normal dispersion.

\begin{figure}[h]
\begin{center}
 \includegraphics[scale=.6]{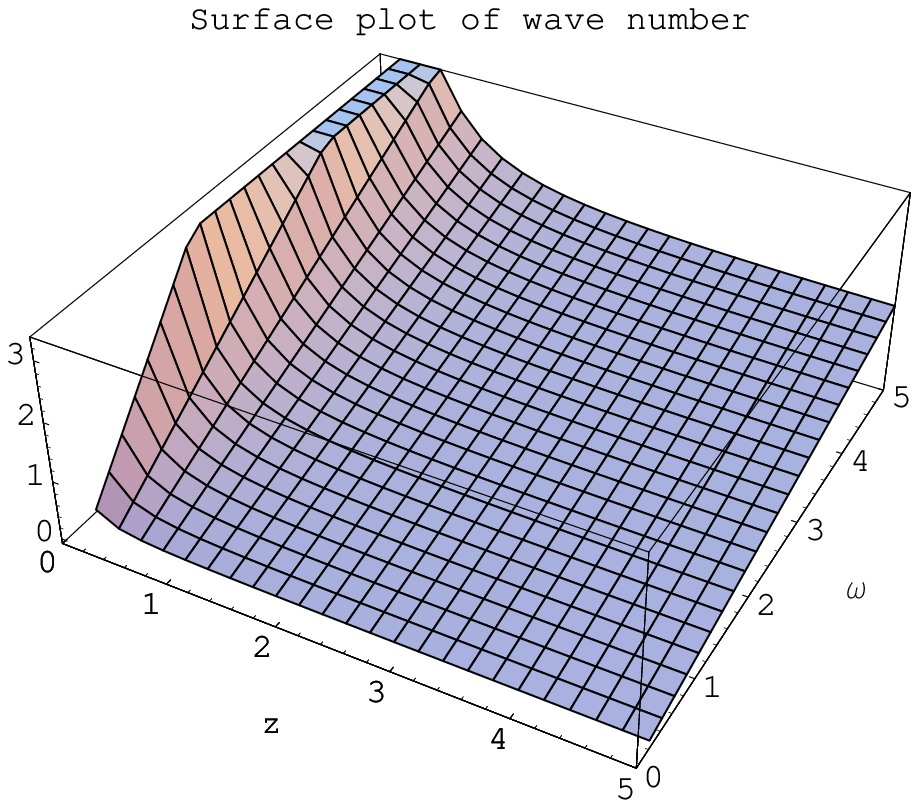}\\
 \includegraphics[scale=.6]{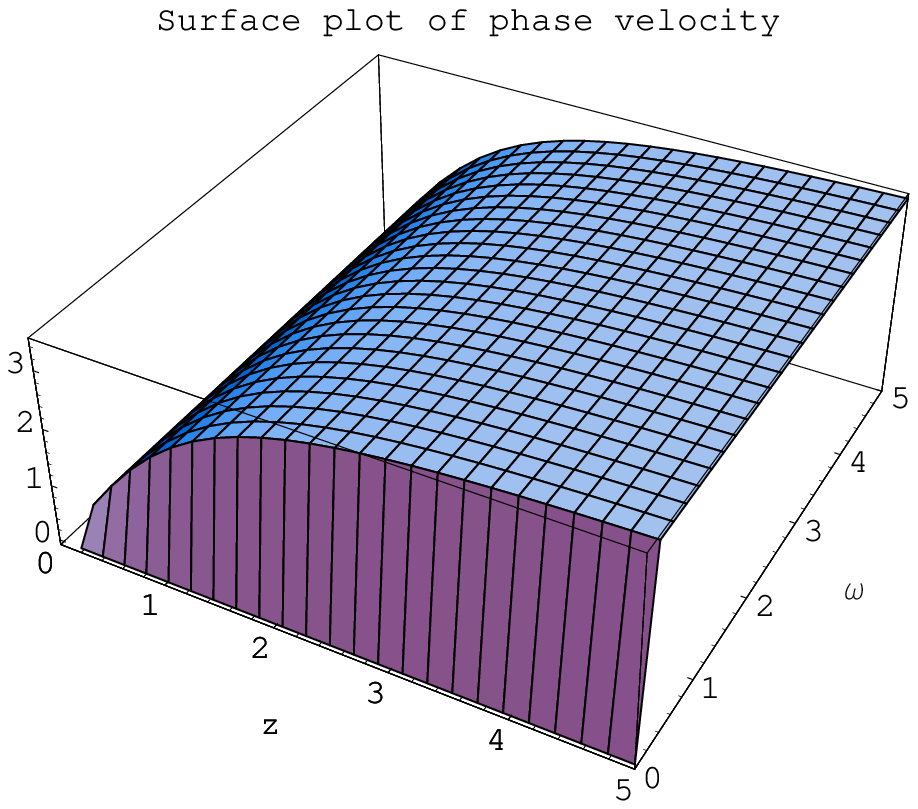}
 \includegraphics[scale=.6]{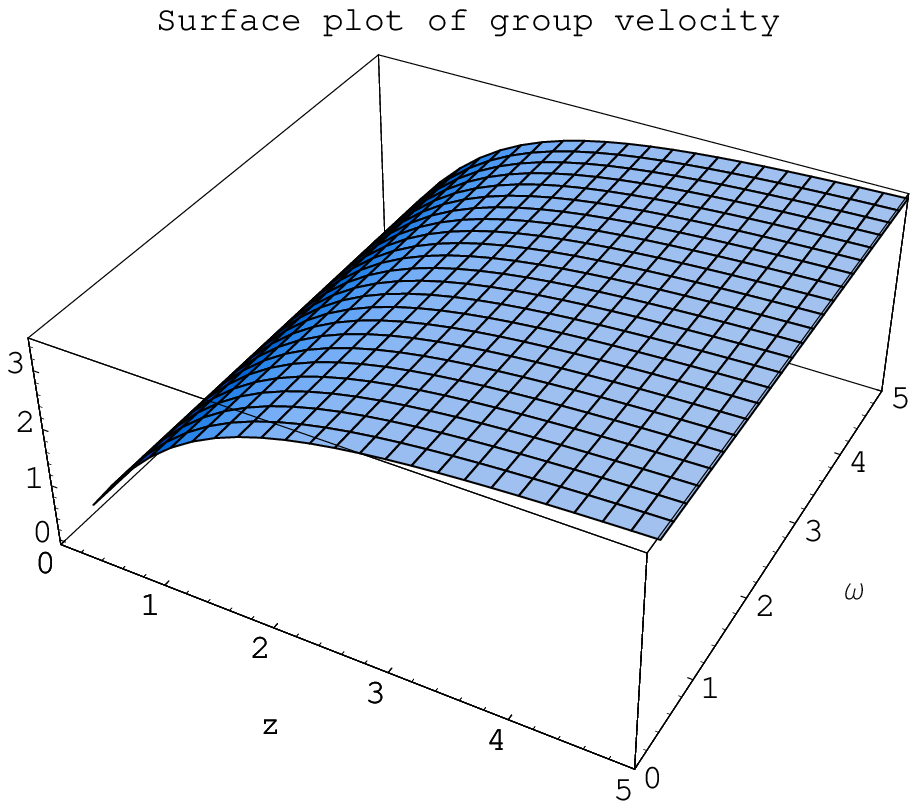}\\
 \includegraphics[scale=.6]{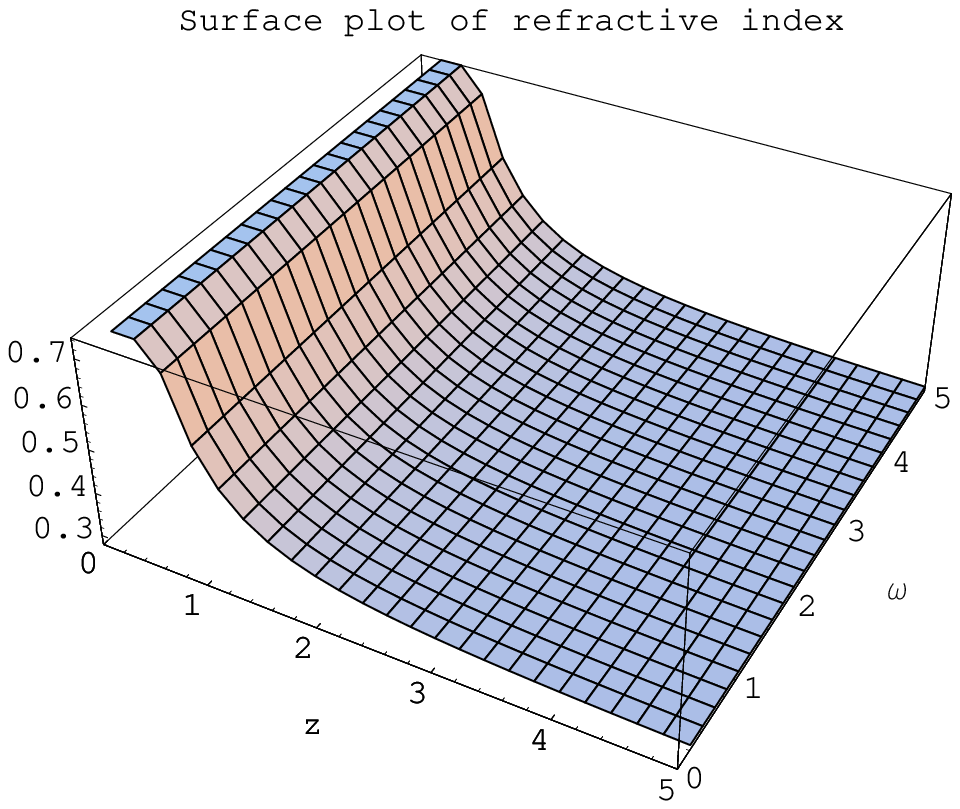}
 \includegraphics[scale=.6]{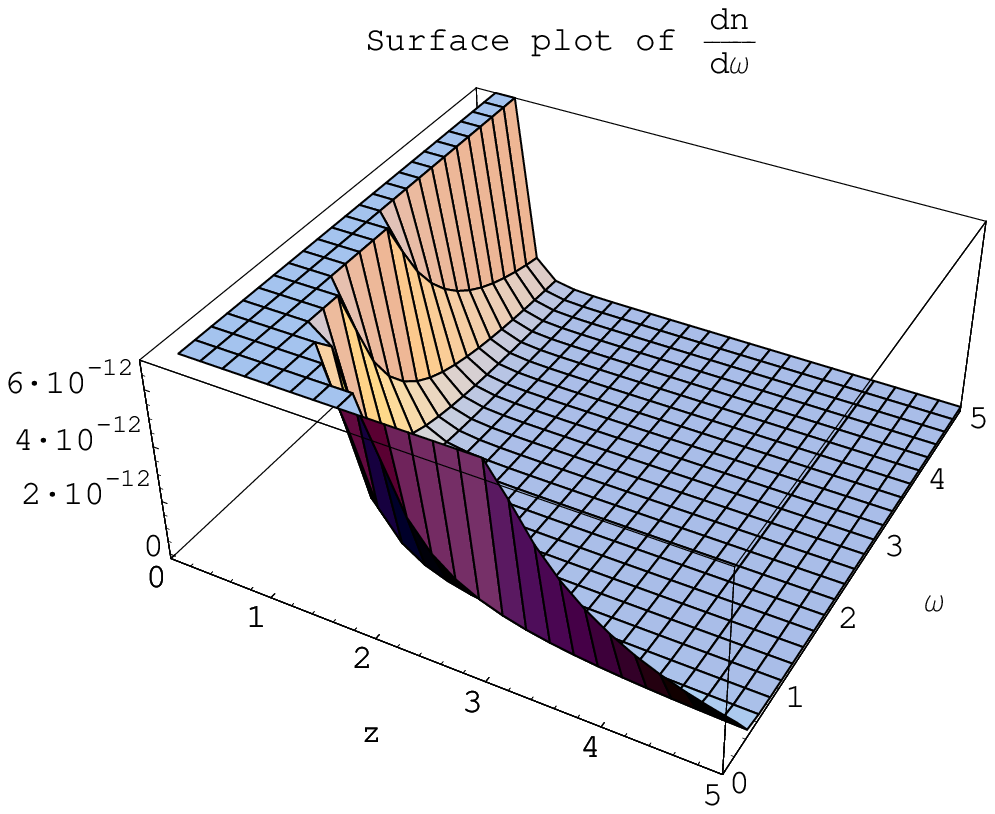}
\end{center}
\caption{The region is not of normal dispersion, phase and group velocities are the same,  $n<1$ and $\frac {dn}{d\omega }>0$.}
\end{figure}

Figure 5 shows that the real waves exist in the region but no wave in the vicinity of the event horizon because of the existence of strong gravitational field there. The wave number increases as angular frequency increases but decreases as $z$ increases. The wave number decreases as we go away from the event horizon and hence damping takes place. For this region refractive index $n<1$ and $\frac{dn}{d\omega}>0$ which implies that the region is of anomalous dispersion.

\begin{figure}[h]
\begin{center}
 \includegraphics[scale=.6]{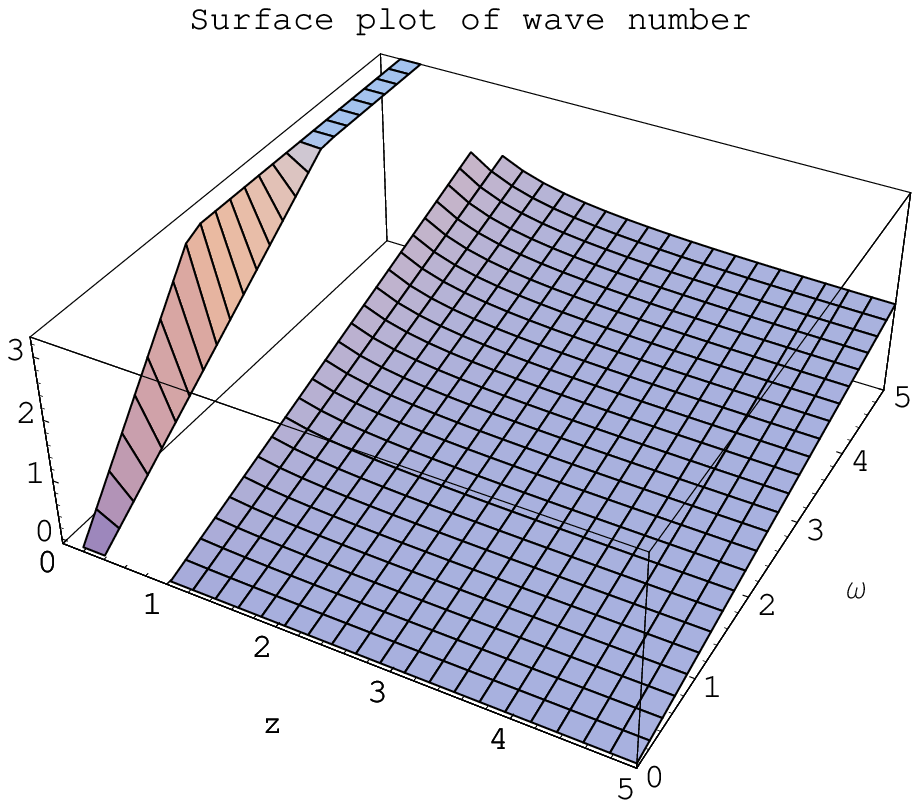}\\
 \includegraphics[scale=.6]{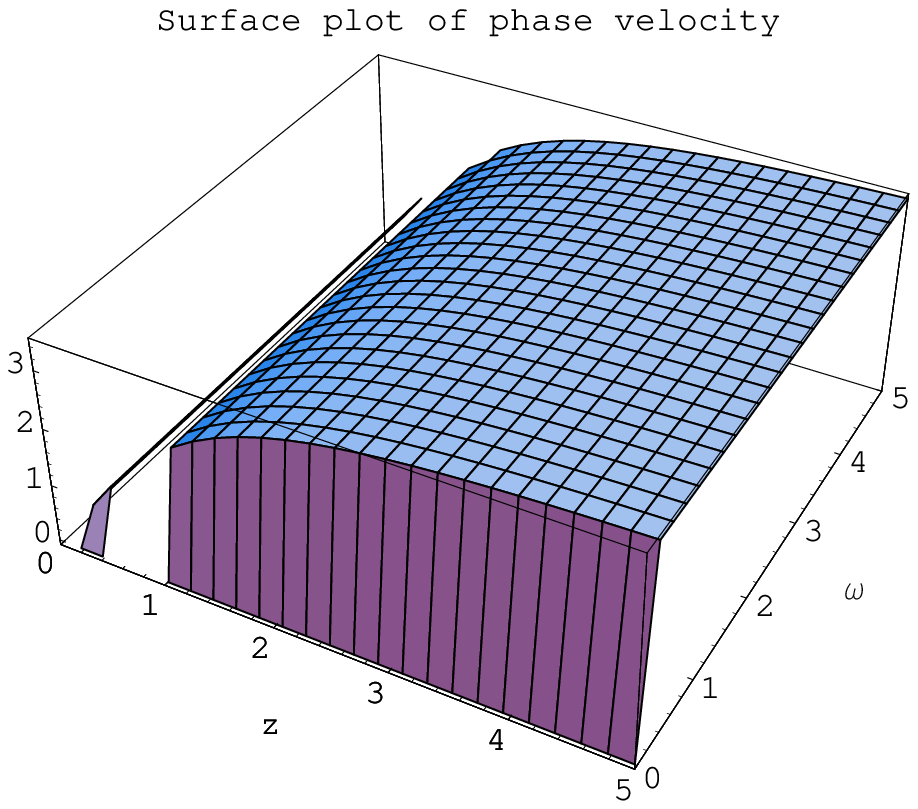}
 \includegraphics[scale=.6]{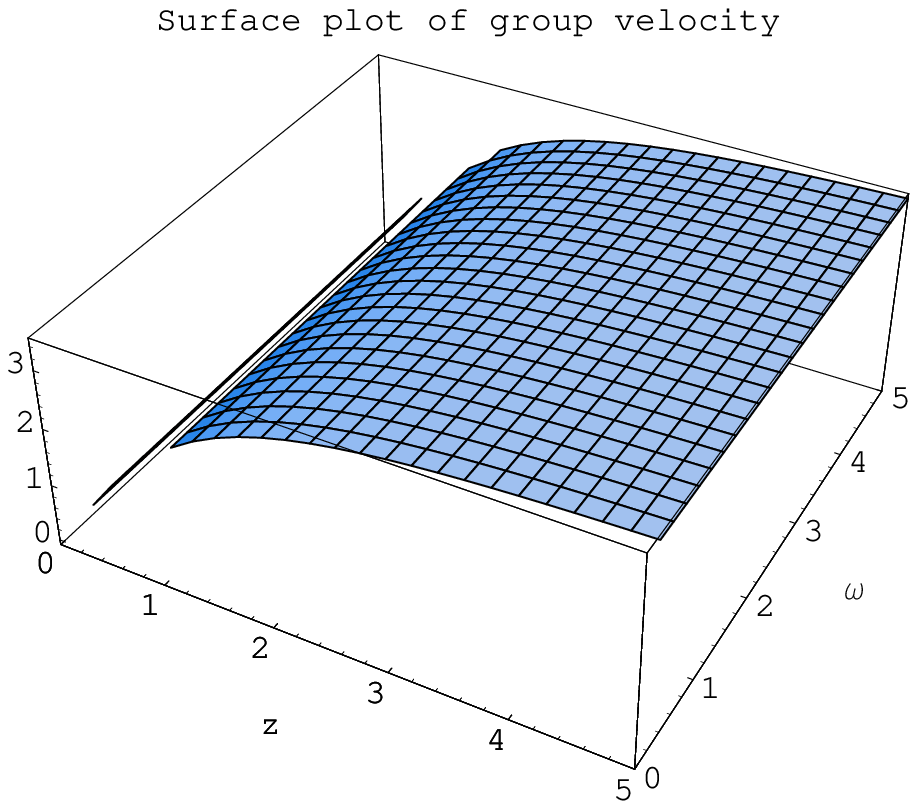}\\
 \includegraphics[scale=.6]{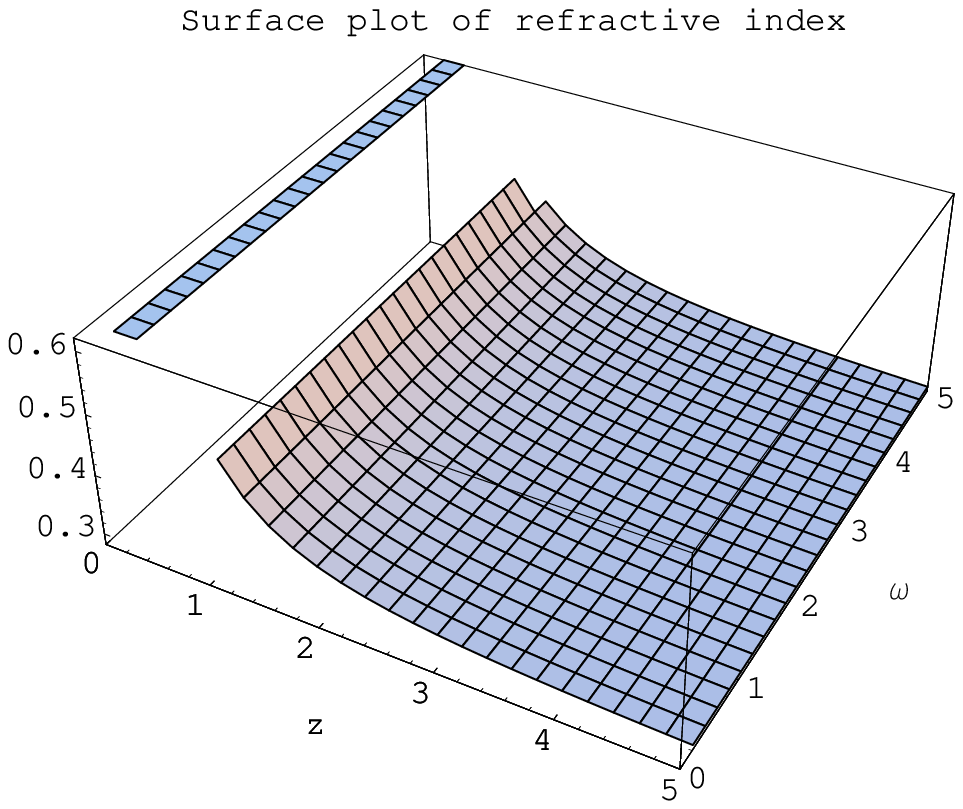}
 \includegraphics[scale=.6]{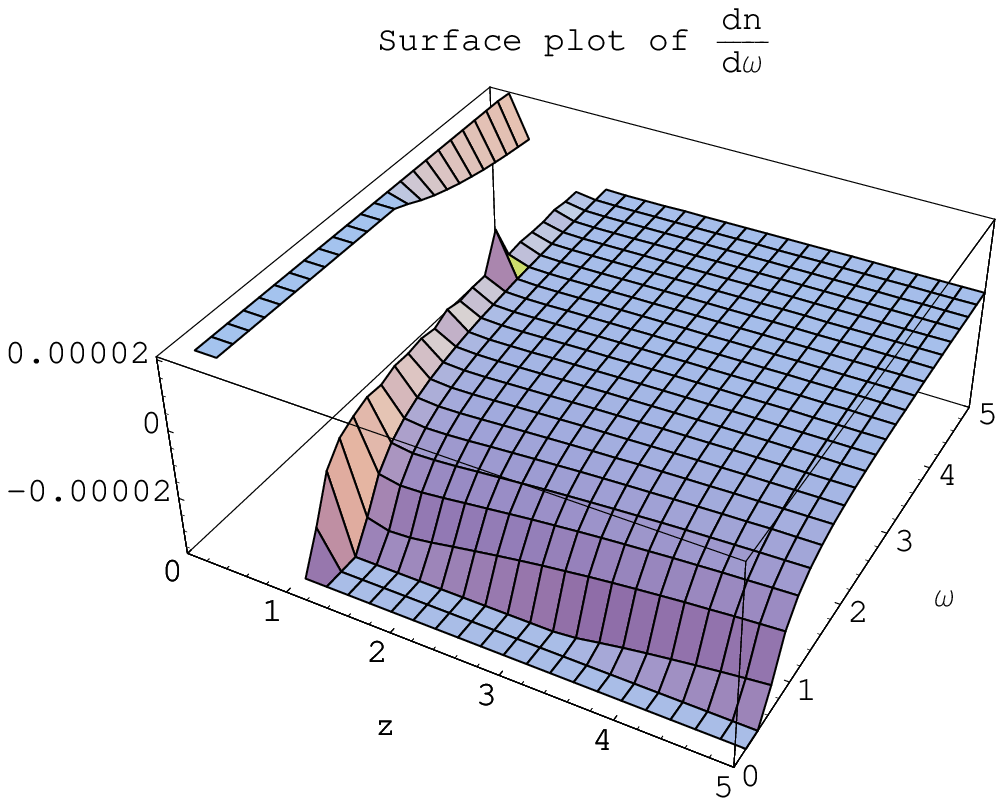}
\end{center}
\caption{The region is not of normal dispersion, phase and group velocities are the same,  $n<1$ and $\frac{dn}{d\omega }<0$ in some region.}
\end{figure}

In Fig. 6, infinite wave number occurs at the horizon, so no wave exists there. The phase and group velocities admit the same pattern. These velocities increase when we depart from horizon. The region has $n<1$ and $\frac{dn}{d\omega}<0$ for some values of $z$ and $\omega$. Hence it is not the case of normal dispersion.

\section{Cold Plasma in Non-rotating Background}\label{sec5}

The magnetosphere has the perturbed flow only along $z$-axis in this background. Hence, FIDO measured fluid four-velocity $\textbf{V}$ is described by
\begin{equation}
\textbf{V}=u(z)\textbf{e}_{\textbf{z}}\label{eq45}.
\end{equation}
The Lorentz factor takes the form
\begin{equation}
\gamma=\frac{1}{\sqrt{1-u^2}}\label{eq46},
\end{equation}

while the FIDO measured magnetic field becomes
\begin{equation}
\textbf{B}=B(z)\textbf{e}_{\textbf{z}}\label{eq47}.
\end{equation}

Then the equations (\ref{eq36})-(\ref{eq41}) reduce to the following form :
\begin{equation}
-\frac{i\omega}{\alpha}c_5=0\label{eq48}
\end {equation}
\begin {equation}
ikc_5=0\label{eq49}
\end{equation}
\begin{equation}
c_1\left(\frac{-i\omega}{\alpha}+iku\right)+c_2\left\{ (1+\gamma^2u^2)ik-(1-2\gamma^2u^2)(1+\gamma^2u^2)\frac {u^\prime}{u}-\frac{i\omega}{\alpha}\gamma^2u\right\}=0\label{eq50}
\end{equation}
\begin{eqnarray}
&&c_1\gamma^2\{a_z+uu^\prime(1+\gamma^2u^2)\}+c_2[\gamma^2 (1+\gamma^2u^2)\left(\frac{-i\omega} {\alpha}+iku\right)\nonumber\\
&&+\gamma^2\{u^\prime(1+\gamma^2u^2)(1+4\gamma^2u^2) +2u\gamma^2a_z\}]=0\label{eq51}.
\end{eqnarray}
From (\ref{eq48}) or (\ref{eq49}), $c_5=0$; hence, there  does not exist any perturbation in magnetic field of the fluid. We also get the same two equations (\ref{eq50}) and (\ref{eq51}) for the non-magnetized plasma. The complex dispersion relation, that follows from (\ref{eq50}) and (\ref{eq51}), is
\begin{eqnarray}
&&u^2\gamma^2(1+u^2\gamma^2)k^2-\{ia_z\gamma^2(-1+u^2\gamma^2) +u(1+u^2\gamma^2)
(3iu^2u^\prime\gamma^2+\frac{2\omega}{\alpha})\}k\nonumber\\
&&-\Big[\gamma^2\{\frac{u^\prime}{u}(a_z+uu^\prime +u^3u^\prime\gamma^2)(1-u^2\gamma^2-2u^4\gamma^4)\nonumber\\
&&-i\{u^\prime+u(a_z+4uu^\prime)\gamma^2+3u^4u^\prime\gamma^4\} \frac{\omega}{\alpha}-(1+u^2\gamma^2)
\frac{\omega^2}{\alpha^2}\}\Big]=0.\label{eq52}
\end{eqnarray}

\begin{figure}[h]
\begin{center}
 \includegraphics[scale=.6]{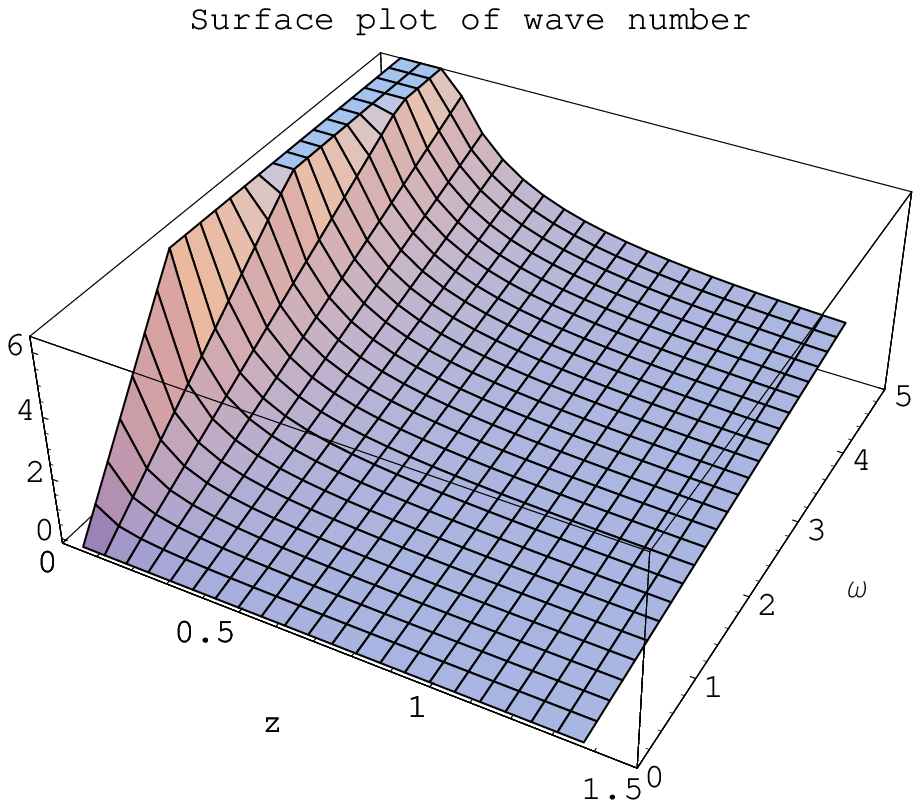}\\
 \includegraphics[scale=.6]{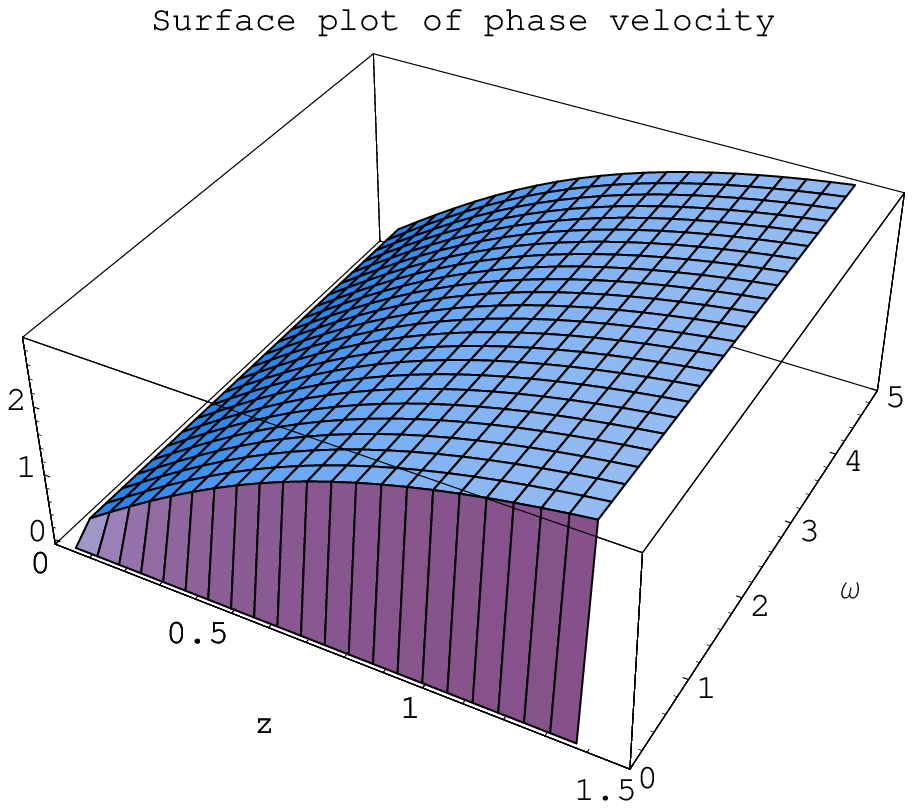}
 \includegraphics[scale=.6]{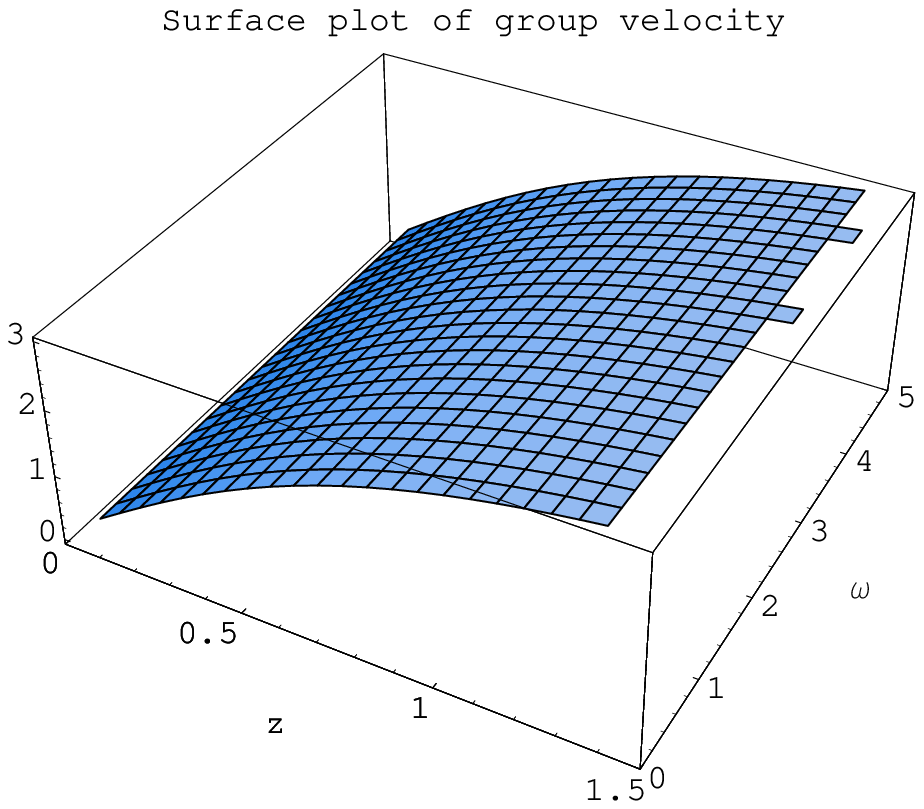}\\
 \includegraphics[scale=.6]{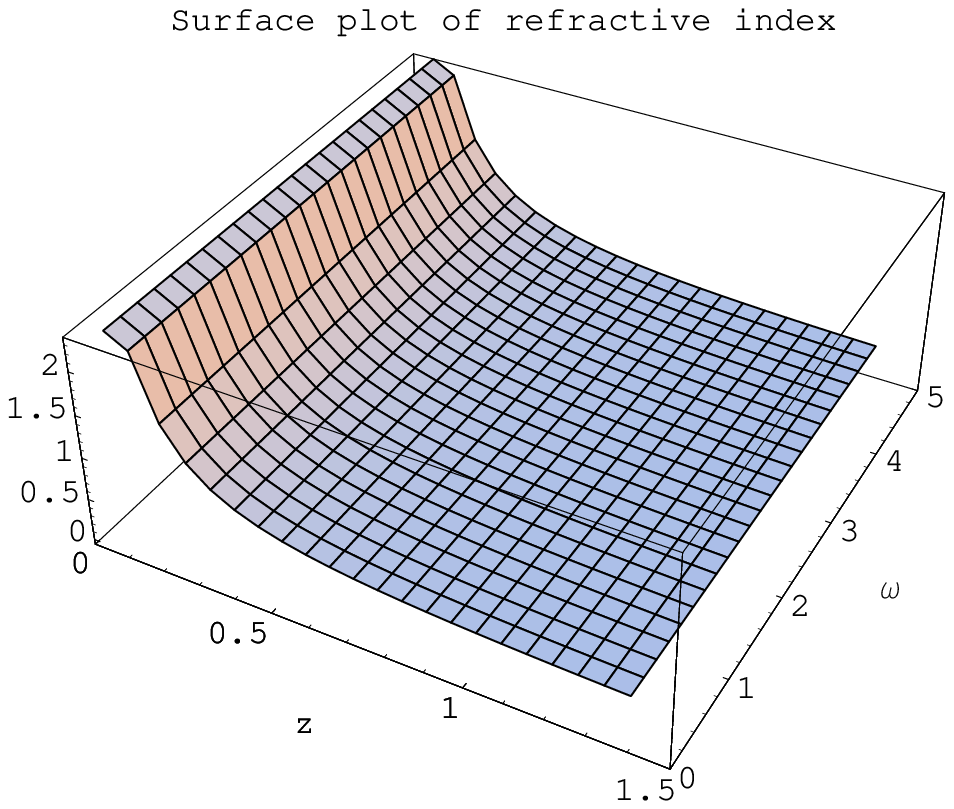}
 \includegraphics[scale=.6]{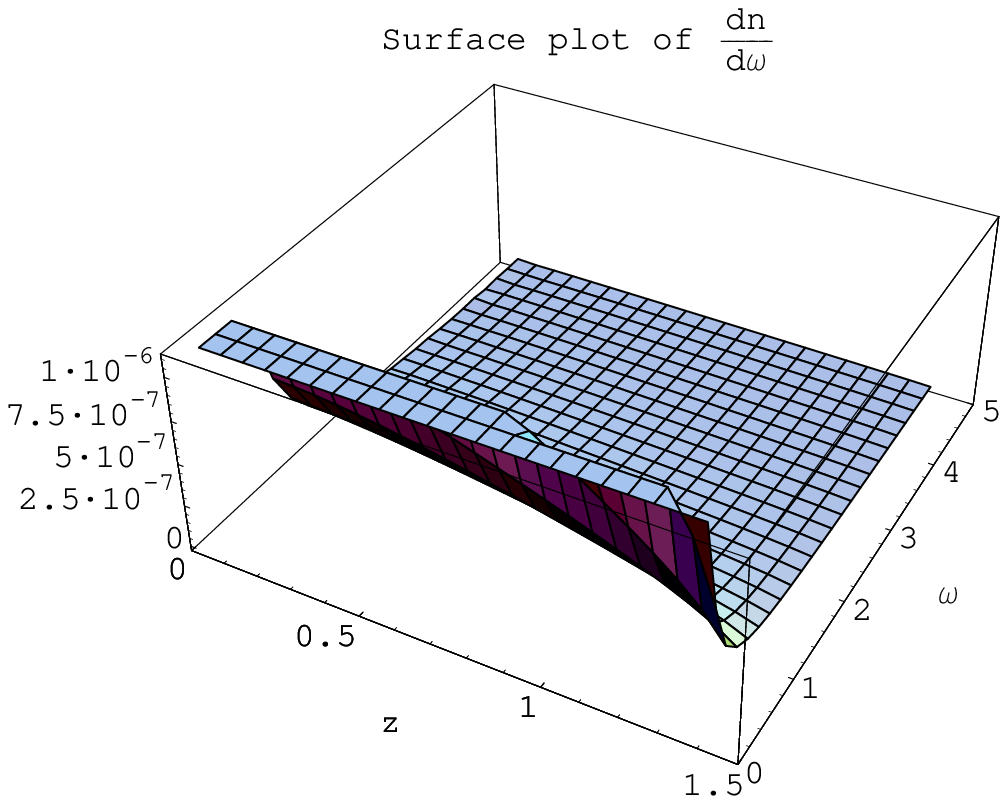}
\end{center}
\caption{The region is of normal dispersion, phase and group velocities are the same, $n>1$ and $\frac{dn}{d\omega}>0$.}
\end{figure}

We investigate the  longitudinal waves propagating parallel to the magnetic field $\textbf{B}$. We analyze the numerical modes for $Q^2/M^2=0.5$. From the mass conservation law we get $u=\frac{1}{\sqrt{z^2+1}}$. The real part of (\ref{eq52}) gives us two values for $k$ in terms of $z$ and $\omega $, shown in Fig. 7 and Fig. 8, while the imaginary part provides one real value for $k$, shown in Fig. 9.

\begin{figure}[h]
\begin{center}
 \includegraphics[scale=.6]{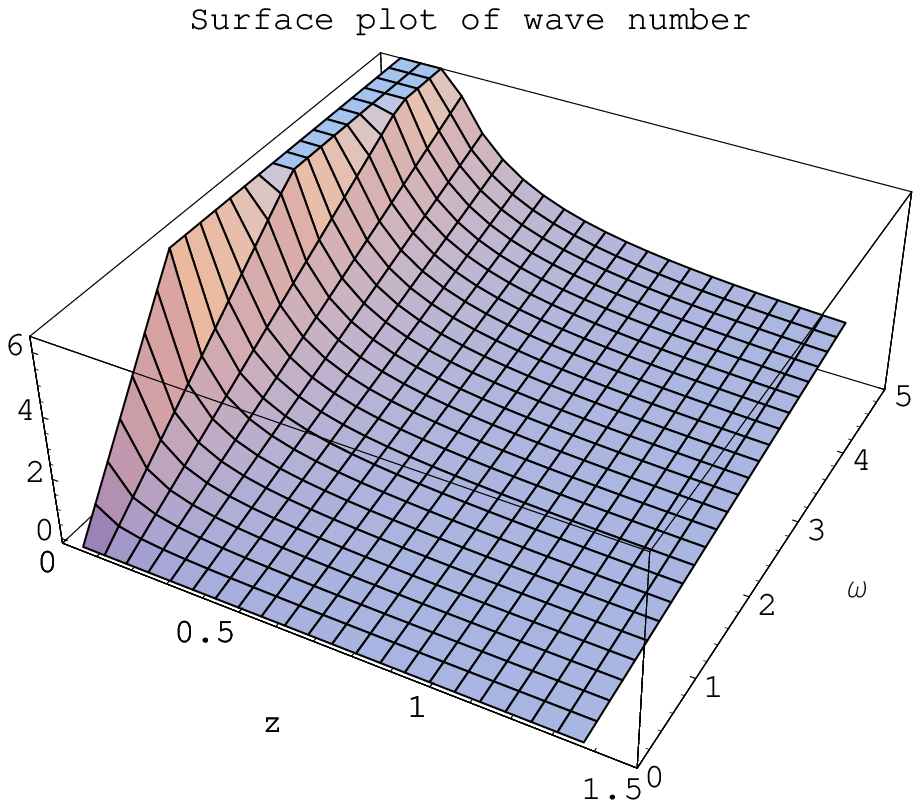}\\
 \includegraphics[scale=.6]{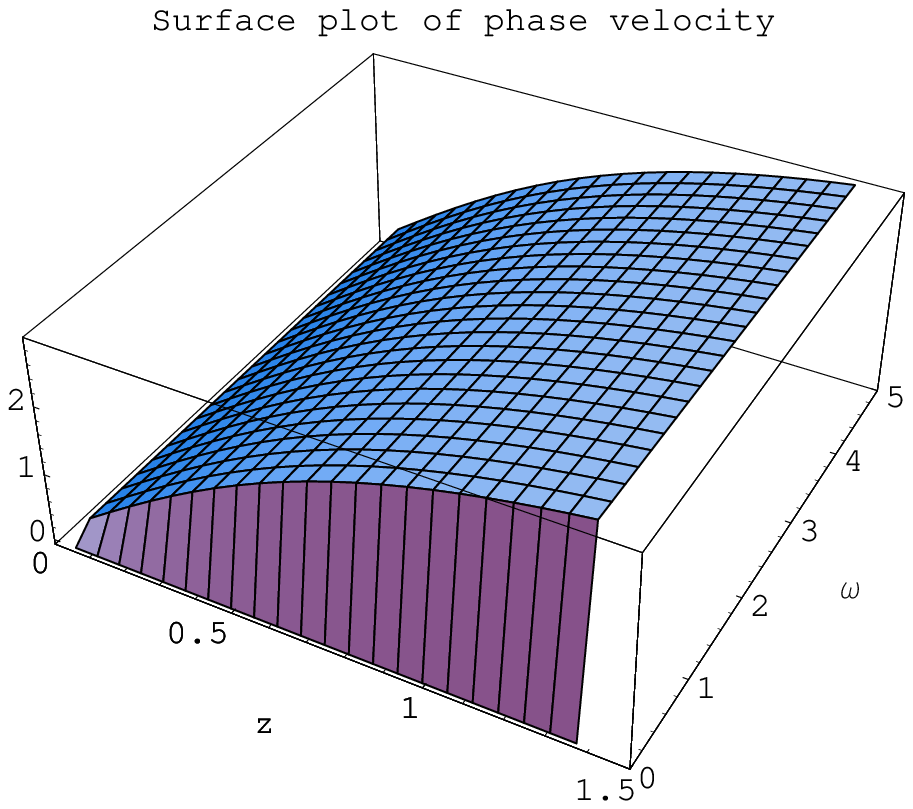}
 \includegraphics[scale=.6]{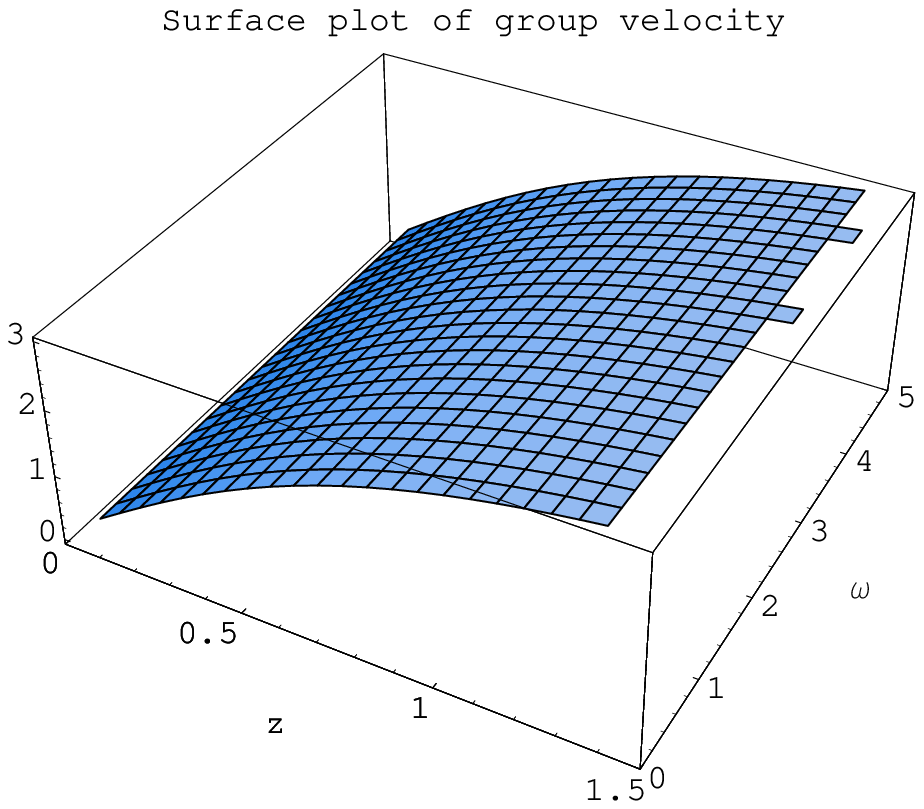}\\
 \includegraphics[scale=.6]{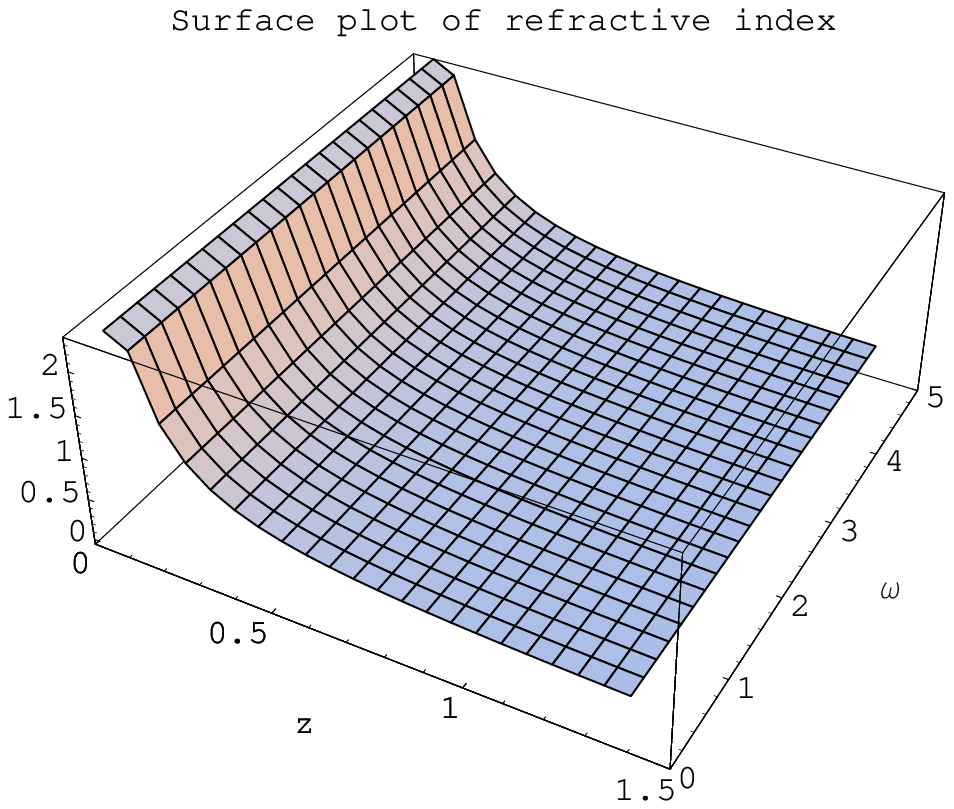}
 \includegraphics[scale=.6]{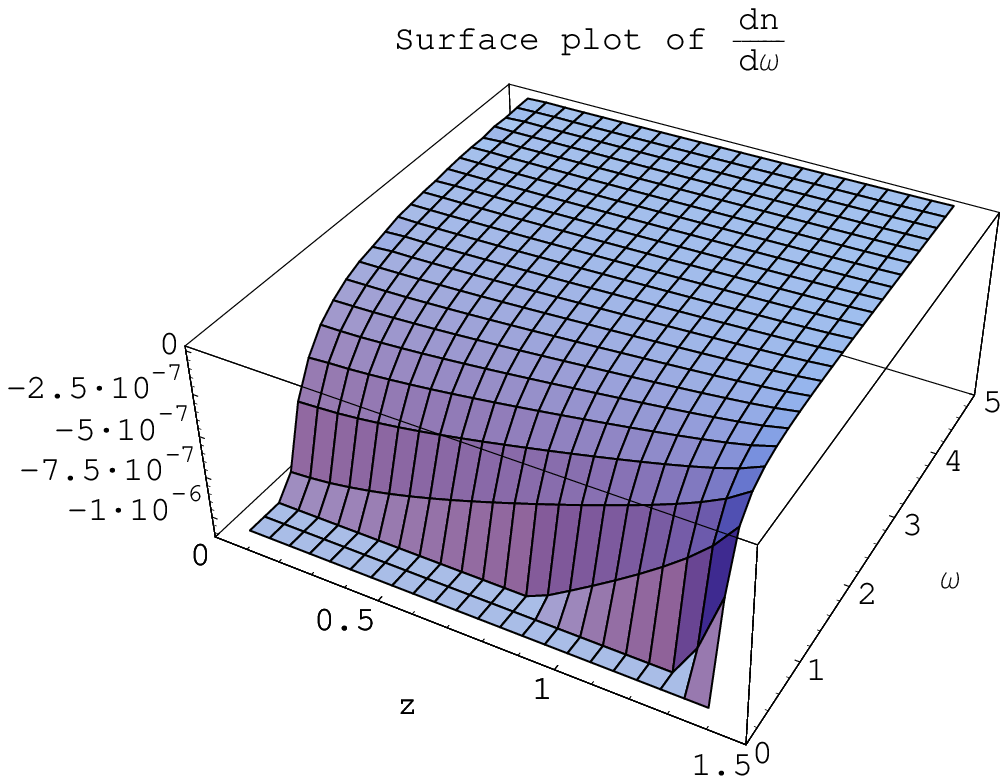}
\end{center}
\caption{The region is not of normal dispersion, phase and group velocities are the same,  $n>1$ but $\frac {dn}{d\omega }<0$.}
\end{figure}

Figure 7 shows that the wave number $k$ decreases as $z$ increases i.e. the waves are growing energy with increase in $\omega$ and decrease in $z$ but damping occurs when $z$ raises. So waves drop energy when we depart from event horizon. At the event horizon of the hole, we see that the wave number becomes infinite which means that the waves disappear due to the effect of immense gravity. The group and phase velocities are same. The refractive index $n>1$ and $\frac{dn}{d\omega}>0$ in this case. Hence the dispersion is normal.

We observe from Fig. 8 that the wave number is infinite at $z=0$ and hence no wave exists there. The wave number decreases as we depart from the event horizon. The waves show damping modes for increasing $z$. The increase in $\omega$ increases $k$. The phase and group velocities have the same behavior. Since $\frac{dn}{d\omega}<0$, the region is not of normal dispersion.

\begin{figure}[h]
\begin{center}
 \includegraphics[scale=.6]{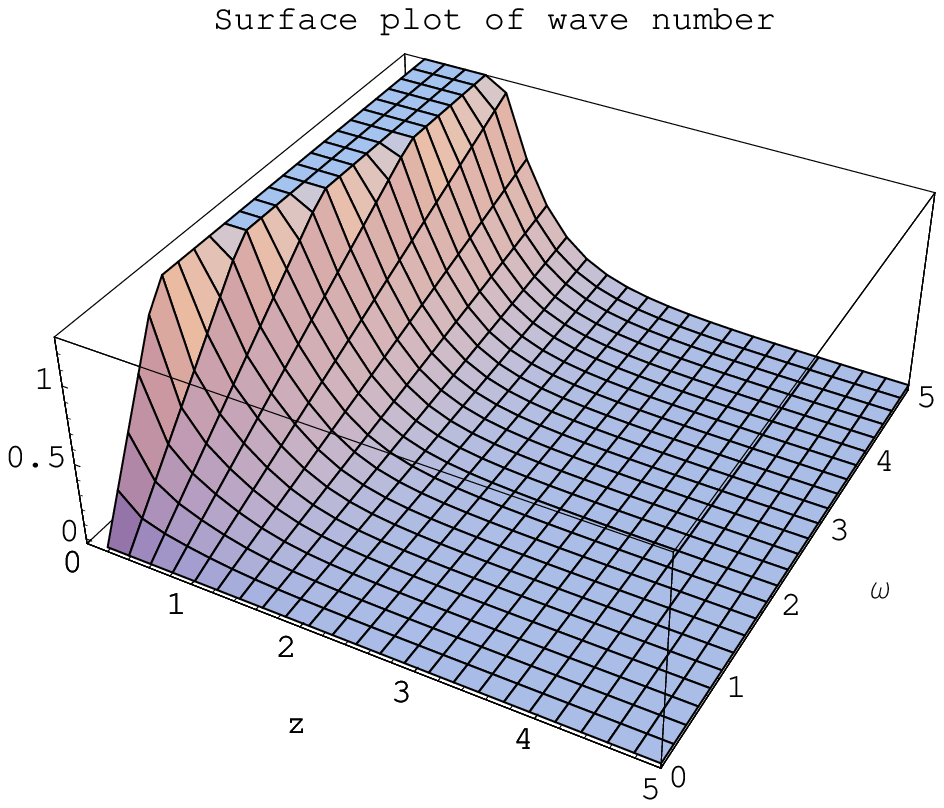}\\
 \includegraphics[scale=.6]{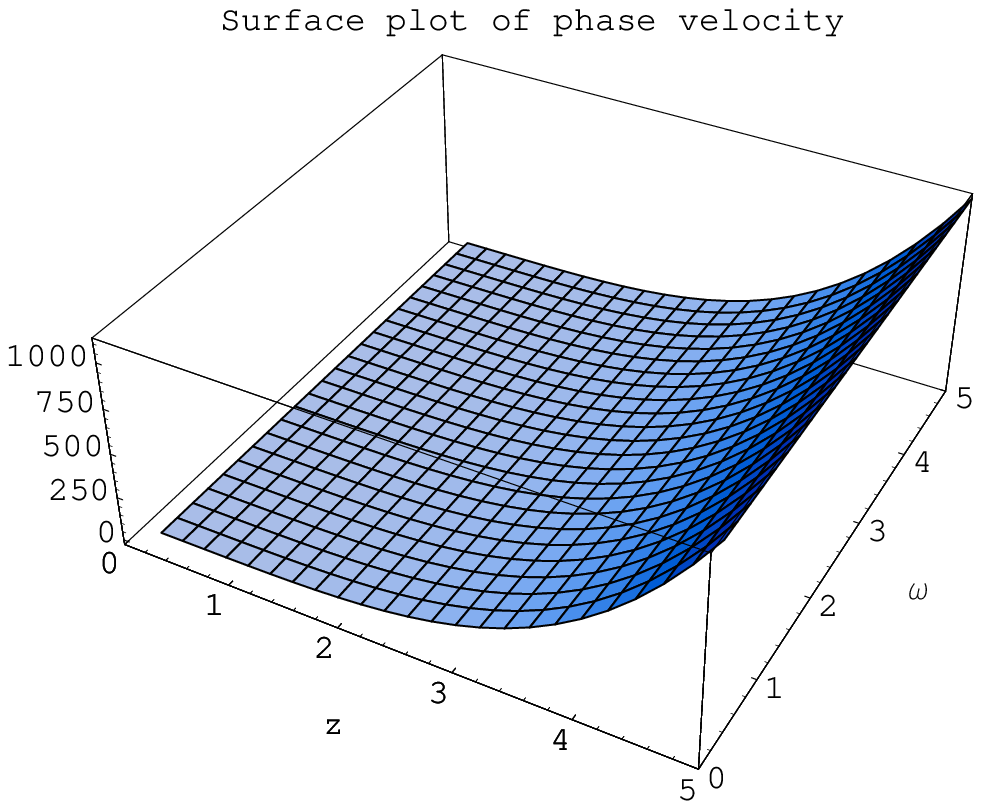}
 \includegraphics[scale=.6]{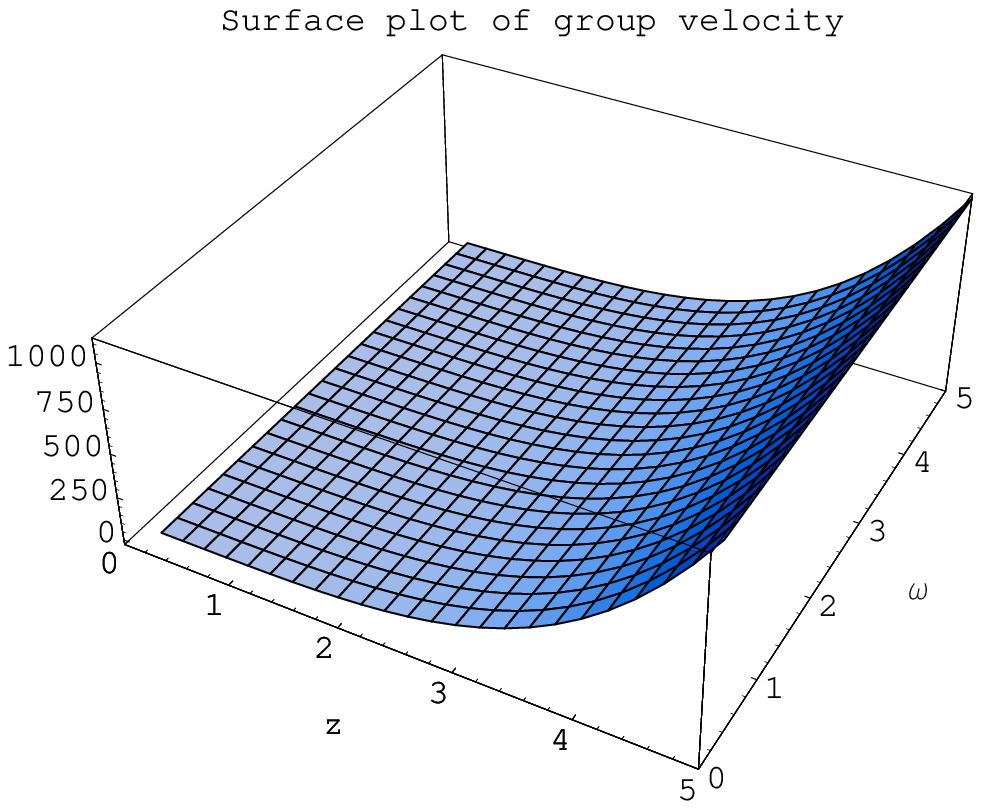}\\
 \includegraphics[scale=.6]{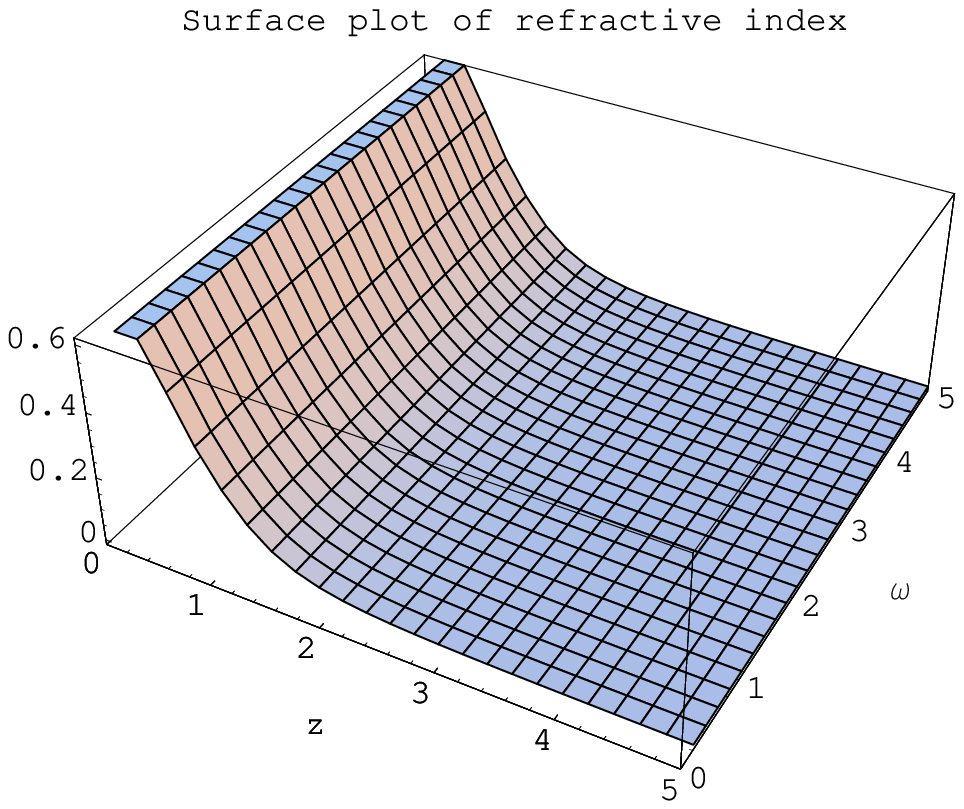}
 \includegraphics[scale=.6]{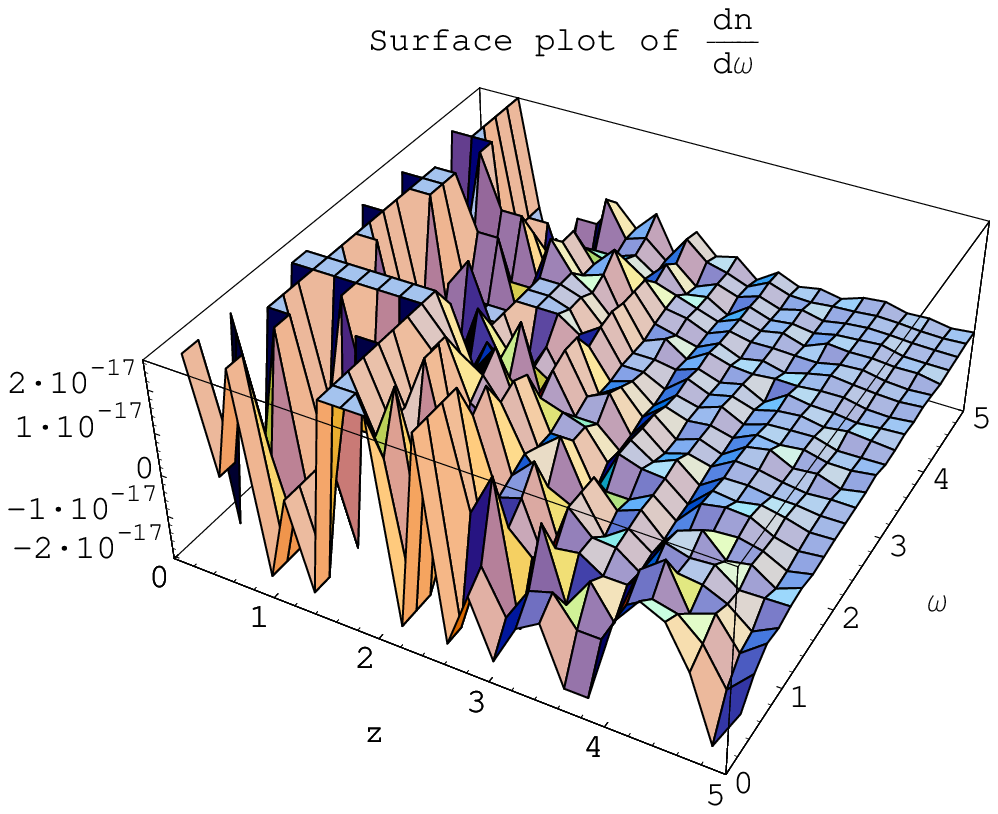}
\end{center}
\caption{The region is not of normal dispersion, phase and group velocities are the same,  $n<1$ and $\frac {dn}{d\omega}\leq0$ in most region.}
\end{figure}

Figure 9 shows that the wave number becomes very large and hence there exists no wave very close to event horizon. The waves are gaining energy with the increase in $\omega$ but losing with the increase in the distance from the event horizon. The surfaces of group and phase velocities are showing the same nature. Since $n<1$ and $\frac{dn}{d\omega}\leq0$ for most region, so the region is not of normal dispersion.

\section{Concluding Remarks}\label{sec6}

Our main concern has been the investigation of the wave properties of cold plasmas in the  RN black-hole's magnetosphere. We have exploited the 3+1 formalism of general relativity, developed by Thorne et al \cite{five,six,seven,eight}, and derived the GRMHD equations in component form by linear perturbation. These equations are then Fourier analyzed to obtain dispersion relations. We have considered both non-rotating and rotating backgrounds (either non-magnetized or magnetized). The properties of plasma waves are analyzed on the basis of the quantities: wave number, phase and group velocities, and refractive index, derived from the dispersion relations. We solve the dispersion relation numerically and present the results graphically.

Our investigation shows that the wave number becomes infinite at the event horizon and consequently, no wave is present there due to immense gravitational field. This indicates that no signal can pass the event horizon or near to it. But when we depart from horizon, the waves lose energy. Wave number is directly proportional to angular frequency and inversely proportional to $z$. Therefore the waves are in damping mode as we go away from the horizon and in growing mode as we approach the horizon.

For a non-rotating background, the magnetospheric fluid disperses normally. Figure 7 shows the normal dispersion, while Fig. 8 and Fig. 9 show dispersions which are not normal. In the case of a rotating non-magnetized background, shown in Fig. 4, Fig. 5, and Fig. 6, we have found cases which are not normally dispersive.

For a rotating magnetized background, we have found a region, shown in Fig. 3, where wave number, phase velocity, group velocity, refractive index, and $\frac{dn}{d\omega}$ are all negative. This implies that the region has all the characteristics of left-handed metamaterials. The group velocity is greater than the phase velocity and $\frac{dn}{d\omega}<0$, which mean that the region is of anomalous dispersion. The dispersion relations provide that the region under discussion is of non-normal dispersion. Hence real signals cannot pass through this region.

Our study shows that the wave number becomes infinite at the event horizon and hence no wave exists there, which supports the well-known point of view that no information can be extracted from a black hole. Mackay et al. \cite{fourty five} found that rotation of a black hole is required for the existence of negative phase velocity propagation and the waves of less angular velocity are evanescent. It is interesting to mention that our analysis shows that negative phase velocity propagates in the rotating background whether the black hole is rotating or non-rotating.

The MHD waves are non-dispersive in the cold plasma, but our analysis, shown graphically, predicts that they are dispersive. This happens because of the formalism used and the equations which are different from the usual MHD equations. Since the 3+1 split of general relativity is used in investigating waves propagating in a plasma influenced by the gravitational field, internal gravity waves which interrupt the MHD waves imply the cases of dispersion in each of the hyperfsurface. The figures of the waves are given in a particular hypersurface of constant time $t$ but not in the whole RN background. Hence our finding is justified locally, not globally.

The result, obtained in this paper, reduces to the case of the Schwarzschild black for $Q=0$, as was obtained in \cite{twenty three}. As described in the introduction, the Reissner-Nordstr\"{o}m solution describes several special charged solutions in special cases, such as, (a) the magnetically charged solution for $Q_e=0$, $0<Q_m<M$, (b) the generic solution for $Q_m=0$, $0<Q_e<M$, and (c) the extremal solution for $Q=M$. Thus the result of this study can be specialized for these interesting solutions by suitably choosing the black hole parameters. The extreme RN black hole is distinguished by its coldness (vanishing Hawking temperature) and its supersymmetry. It occupies a special position among the black-hole solutions to the Einstein or Einstein-Maxwell equations because of its complete stability with respect to both classical and quantum processes permitting its interpretation as a soliton \cite{fourty six,fourty seven}. The extremal spacetime is also special in admitting supersymmetry in the context of $N=2$ supergravity \cite{fourty seven,fourty eight,fourty nine,fifty,fifty one,fifty two}. Thus, aspects of the Reissner-Nordstr\"{o}m solution might be of interest in a broader context.

In view of the above reasons, our study of the dispersion relation for the cold plasma near the event horizon of the RN black hole is well motivating. The results of this paper can be extended to the RN spacetime generalized with a cosmological parameter. This type of extension may be interesting from the point of view of an inflationary scenario of early universe.

\vspace{1.0cm}

\noindent
{\large\bf Acknowledgement}\\
MHA wishes to acknowledge Associateship of ICTP, Trieste, Italy. MKH is thankful to SUST, Sylhet, Bangladesh for granting leave during this work.

\end{document}